\documentclass{ws-ijcm}

\usepackage{multirow}

\begin{document}

%
\catchline{}{}{}{}{}
%

\markboth{R. A. Amaro Junior, L.-Y. Cheng \textit{\&} S. K. Buruchenko}
{A comparison between WCSPH and MPS methods for 3D dam-break flows}

\title{A COMPARISON BETWEEN WEAKLY-COMPRESSIBLE SMOOTHED PARTICLE HYDRODYNAMICS (WCSPH) AND MOVING PARTICLE SEMI-IMPLICIT (MPS) METHODS FOR 3D DAM-BREAK FLOWS}

\author{RUBENS A. AMARO JUNIOR\footnote{Corresponding author.}$^{\:\:,\dagger}$, LIANG-YEE CHENG$^{\mathsection,\dagger}$ and SERGEI K. BURUCHENKO$^{\mathparagraph,\ddagger}$}

\address{$^\dagger$Department of Construction Engineering, University of São Paulo, Av. Prof. Almeida Prado, trav. 2, 83 - Cidade Universitária\\
São Paulo, SP, 05508-900, Brazil\\
$^\ddagger$South Ural State University, Lenin Av., 76\\
Chelyabinsk, 454080, Russia Federation\\
\email{$^*$rubens.amaro@usp.br\\
$^\mathsection$cheng.yee@usp.br\\
$^\mathparagraph$buruchenkosk@susu.ru}
}





\maketitle

\begin{history}
\received{(Day Month Year)}
\revised{(Day Month Year)}
\end{history}

\begin{abstract}
Lagrangian particle-based methods have opened new perspectives for the investigation of complex problems with large free-surface deformation. Some well-known particle-based methods adopted to solve non-linear hydrodynamics problems are the smoothed particle hydrodynamics (SPH) and the moving particle semi-implicit (MPS). Both methods modeled the continuum by a system of Lagrangian particles (points) but adopting distinct approaches for the numerical operators, pressure calculation, and boundary conditions. Despite the ability of the particle-based methods in modeling highly nonlinear hydrodynamics, some shortcomings, such as unstable pressure computation and high computational cost remains. In order to assess the performance of these two methods, the weaklycompressible SPH (WCSPH) parallel solver, DualSPHysics, and an in-house incompressible MPS solver are adopted in this work. Two test cases consisting of three-dimensional (3D) dam-break problems are simulated, and wave heights, pressures and forces are compared with available experimental data. The influence of the artificial viscosity on the accuracy of WCSPH is investigated. Computational times of both solvers are also compared. Finally, the relative benefits of the methods for solving free-surface problems are discussed, therefore providing directions of their applicability
\end{abstract}


\section{Introduction}
Particle-based methods are very effective for the simulation of hydrodynamics problems involving free surface, fragmentation and merging, large deformation, complex-shaped bodies and moving boundaries. Among those methods, smoothed particle hydrodynamics (SPH) and moving particle semi-implicit (MPS) have opened new perspectives to solve non-linear hydrodynamics problems. The SPH was first introduced in astrophysics by \citeauthor{gingold77} \shortcite{gingold77} and \citeauthor{lucy77} \shortcite{lucy77}, whereas the MPS was proposed by \citeauthor{koshizuka96} \shortcite{koshizuka96} for the simulation of incompressible free-surface flow.

SPH and MPS are very similar since they are purely meshfree, use particles (points) to represent the continuum and approximate the governing equations in strong form, based on local averaged summations \cite{liu15}. Also, in both methods the particles interact with each other within a compact support. Currently, both SPH and MPS adopt explicit and semi-implicit algorithms, as well as share many similarities, e.g., numerical techniques used to free surface detection, solid boundary conditions, multi-phase interface, surface tension models, and many others. The main differences between them remain in the formulation of the spatial discretization. In the MPS, differential operator models are approximated using a local weighted difference of the variables without taking the gradient of a kernel function. On the other hand, in the SPH, a global distribution of a variable is constructed by a superposition of kernels \cite{koshizuka18} and the differential operator models are approximated using averaged summation based on the kernel gradient. Nonetheless, once can make use of equivalence between the MPS weighting function and the SPH kernel to relate the SPH and MPS operators, as established in \citeauthor{souto13} \shortcite{souto13}.

Despite the ability of the particle-based methods in modeling highly distorted free-surface flow, wave breaking and flow separation and merging, some shortcomings of these methods, such as unstable pressure computation and high computational cost remains. Different approaches were proposed to achieve more stable computation, e.g., enhancement of regular particle distributions by adopting particle regularization techniques, modified and/or high-order differential operators, formulations for the source term of PPE combining incompressibility conditions or introducing high-order source terms and accurate boundary treatment techniques used in free-surface detection, solid boundary and multiphase interface. In that context, one may refer to \cite{gotoh16,gotoh18,liu19,ye19} wherein there are more referred related works of both weakly-compressible and incompressible particle-based methods. Concerning the simulation of large-scale practical engineering problems in a reasonable processing time, studies on parallel processing in central processing units (CPUs) \cite{fernandes15,guo18} or graphics processing units (GPUs) \cite{hori11,chow18} have been reported.

In order to identify the performance of each formulation under specific applications, several works compared weakly-compressible particle-based methods, in which a fully explicit algorithm and an equation of state assuming a limited-compressible flow is adopted, and incompressible ones based on semi-implicit algorithm with the solution of the pressure Poisson equation (PPE) to model the incompressible flow. \citeauthor{lee08} \shortcite{lee08} presented comparisons between the classical weakly-compressible SPH (WCSPH) and the incompressible SPH (ISPH) \cite{cummins99} methods for two-dimensional (2D) problems. According their work, the ISPH yielded much more reliable results than WCSPH, shown smoother velocity and pressure fields. \citeauthor{hughes10} \shortcite{hughes10} also compared WCSPH and ISPH for 2D dam-break problems and regular water waves impacting against a vertical wall. Moving least-squares (MLS) or Shephard filtering density renormalization methods and specific boundary condition formulations were used. As a result, they concluded that in the optimum configuration, WCSPH performs at least as well as ISPH, and in some respects clearly performs better. \citeauthor{rafiee12} \shortcite{rafiee12} compared the accuracy, stability and efficiency of the SPH by using acoustic Riemann solver, named Godunov SPH (GSPH) method, against the WCSPH and ISPH techniques for 2D non-linear high Reynolds number sloshing flows. Overestimated and highly oscillatory time variation of impact pressure was computed when WCSPH was used. The ISPH produced more accurate pressure results, although resulted in a more scattered distribution of particles, and the computational time was reduced since much larger time step is allowed. More accurate, stable and reliable results were computed by using the GSPH but resulting in substantially large computational costs. In \citeauthor{hashimoto13} \shortcite{hashimoto13}, comparisons of experiment data and the WCSPH and MPS results were performed on forced roll tests of a 2D damaged car ferry. Both numerical results were in good agreement with the model experiment when the entrapped air effect was relatively small. However, the agreement become worse when air entrapped occurred. Based on the experimental results for a collapsing water column with a rigid obstacle and on wet bed, \citeauthor{abdelrazek14} \shortcite{abdelrazek14} compared 2D dam-break numerical results obtained by WCSPH and standard MPS, i.e., without recent improvements. As a result, more stable particles distributions were obtained by WCSPH due its artificial viscosity, along with smoother predicted free surface and pressure time series. Despite the better performance of the WCSPH, the authors highlighted the independence of MPS regarding the parameters used in WCSPH, which usually require appropriate tuning. In \citeauthor{zheng14} \shortcite{zheng14}, the accuracy and efficiency of WCSPH and ISPH are assessed and the results compared with the experimental data from 2D dam-break and sloshing flows. From their results, ISPH gives smoother and more reasonable pressures than that computed by WCSPH. Since the WCSPH requires small time steps, they showed that the efficiency of ISPH can be higher than that of WCSPH. In \citeauthor{bakti16} \shortcite{bakti16}, experiments of 2D dam breaking and harmonically oscillated sloshing tank were adopted as references to compare the results obtained by WCSPH and MPS. A free-decay test case also was studied. In the dam-break case, a new relation between the artificial viscosity coefficient of the WCSPH and the kinematic viscosity of the MPS was obtained. Both methods showed reasonably good agreement with the experimental results. However, a small gap between the fluid and the wall was shown in the WCSPH simulation due to the formulation of the impenetrable wall boundary condition. The authors also pointed out that the use of WCSPH, without a careful treatment, can significantly overestimate the impact pressure. Moreover, WCSPH seems to be more efficient computationally but without solving PPE is generally less accurate. \citeauthor{jian16} \shortcite{jian16} used WCSPH and ISPH to simulate 2D and three-dimensional (3D) dam-break flows impacting either a fixed or a movable structure and compared their accuracy and efficiency. Overall, they founded that the ISPH is slightly superior to the WCSPH approach since more stable particle motion and pressure distribution were achieved with a reasonable computational time. Accuracy, stability and efficiency of WCSPH and ISPH methods are compared in simulating 2D dam-break flow in \citeauthor{akbari18} \shortcite{akbari18}. In his work, more stable and accurate pressure field is computed by using ISPH, however its original version using the divergence-free velocity source term suffered from density error accumulation. In addition, it is showed that ISPH is more efficient computationally than WCSPH, even though the computational time of ISPH is more dependent to the number of particles because a system of linear equations should be solved in each step of the solution.

From the previous studies, the strong or weak points of each method depend on the phenomena being investigated, the required output, the available computational resources and the improved variants adopted. The aim of this paper is to explore the merits of WCSPH and MPS in simulations of transient 3D free-surface flow problems, therefore providing additional results for the discussions towards the applicability of both methods. Here, the open source DualSPHysics \cite{crespo15} based on the WCSPH and an in-house program based on MPS were adopted for the simulations. Two test variants of 3D dam-break problems \cite{kleefsman05,aureli15} were considered and the computed wave heights, pressures and forces were compared with available experimental data. As previously discussed and investigated in \citeauthor{padova14} \shortcite{padova14} and for the finite particle method (FPM) in \citeauthor{he20} \shortcite{he20}, the artificial viscosity coefficient plays an important role in breaking processes such as, wave splashing, plunging jet penetration depth, and wave breaking. In such context, the influence of the artificial viscosity on the accuracy of WCSPH is also analyzed herein. The computational times for each method are also presented.

\section{Governing equations and numerical methods}
In this section, the numerical methods are described briefly.

\subsection{Governing equations}
The governing equations of incompressible flow of viscous Newtonian fluids are expressed by the conservation laws of mass and momentum (in a Lagrangian frame of reference):

\begin{equation}
\frac{\mathrm{D}\rho}{\mathrm{D}t}=\rho\nabla\cdot\mathbf{u}=0\,,
\label{eq:mass}
\end{equation}

\begin{equation}
\frac{\mathrm{D}\mathbf{u}}{\mathrm{D}t}=-\frac{\nabla P}{\rho} + \nu\nabla^2\mathbf{u} + \mathbf{f}\,,
\label{eq:momentum}
\end{equation}

where $\rho$ is the fluid density, $\mathbf{u}$ is the velocity vector, $P$ is the pressure, $\nu$ is the kinematic viscosity and $\mathbf{f}$ is vector of the external body force per unit mass.

\subsection{Weakly-compressible smoothed particle hydrodynamics (WCSPH)}
WCSPH is defined as a method for obtaining approximate numerical solutions of fluid dynamics equations by replacing the fluid with a set of particles. The physical properties, for example the time derivatives of density and velocity on each particle are obtained summing weighted contributions from the neighboring particles. The method uses discrete approximations for interpolation integrals to transform differential equations of fluid dynamics into particle summations. For this purpose, weighting functions called smoothing kernels are employed. A complete review on standard WCSPH can be found at \citeauthor{monaghan05} \shortcite{monaghan05}. The artificial viscosity scheme, proposed by \citeauthor{monaghan92} \shortcite{monaghan92}, is adopted here to solve the discretized momentum equation (\ref{eq:momentum}) at the particle $i$ as follow:

\begin{equation}
\frac{\mathrm{D}\mathbf{u}_i}{\mathrm{D}t} = - \sum_{j \in \mathbb{P}_i} \left[m_j \left( \frac{P_i + P_j}{\rho_i\rho_j} + \Pi_{ij} \right)\nabla_i W_{ij} \right] + \mathbf{f}_i\,,
\label{eq:sphu}
\end{equation}

where $m$ is the mass, and $\mathbb{P}_i$ is the set of all the neighboring particles $j$ for a given particle $i$. The viscosity term $\Pi_{ij}$ is given by:

\begin{equation}
\Pi_{ij} = \left\{
\begin{array}{cr}
-\frac{\alpha \overline{c}_{ij} \mu_{ij}}{\overline{\rho}_{ij}} & \:\:\: \mathbf{u}_{ij} \cdot \mathbf{r}_{ij} < 0 \\
0 & \:\:\: \mathbf{u}_{ij} \cdot \mathbf{r}_{ij} \geq 0
\end{array} \right. \,,
\label{eq:pi}
\end{equation}

with

\begin{align}
\overline{c}_{ij} &= \frac{c_i + c_j}{2} \nonumber \, ,\\
\overline{\rho}_{ij} &= \frac{\rho_i + \rho_j}{2} \nonumber \, ,\\
\mu_{ij} &= h \mathbf{u}_{ij} \cdot \frac{\mathbf{r}_{ij}}{\vert\mathbf{r}_{ij}\vert^2 + \eta^2 }\,, \:\:\: \eta^2 = 0.01 h^2\,,
\label{eq:cpmu}
\end{align}

where $\mathbf{r}_{ij} = \mathbf{r}_{j} - \mathbf{r}_{i}$ and $\mathbf{u}_{ij} = \mathbf{u}_{j} - \mathbf{u}_{i}$, with $\mathbf{r}$ being the particle position. The artificial viscosity $\alpha$ is a parameter that needs to be tuned in order to introduce the proper dissipation. $c$ is the artificial speed of sound. The smoothing length $h = k_{SPH} l^0$ controls the size of the area around one particle in which neighboring particles are considered, with $k_{SPH}$ being a constant and $l^0$ the initial distance between two adjacent particles.
The symbol $W_{ij}$ indicates the WCSPH kernel function, which is used here to “weight” the particle interactions. The kernel is expressed as a function of the dimensionless distance between particles $q = \frac{\vert r_{ij} \vert}{h}$, and can be obtained by the Wendland function \cite{wendland95}:

\begin{equation}
W_{ij} = \varpi_D \left( 1 - \frac{q}{2} \right)^4 \left( 2q + 1 \right) \:\:\: 0 \leq q \leq 2 \,,
\label{eq:sphw}
\end{equation}

where $\varpi_D = \frac{7}{4} \pi h^2$ in 2D and $\varpi_D = \frac{21}{16} \pi h^3$ in 3D.

The $\delta$-SPH variant \cite{molteni09}, which introduces a proper artificial diffusive term, is applied to the continuity equation, see Eq.~(\ref{eq:mass}), in order to suppress the spurious numerical high-frequency oscillations that generally affect the pressure field of the WCSPH scheme. Here, the continuity equation is discretized as:

\begin{equation}
\frac{\mathrm{D}\rho_i}{\mathrm{D}t} = \sum_{j \in \mathbb{P}_i} \left( m_j \mathbf{u}_{ij} \cdot \nabla_i W_{ij} \right) + 2 \delta h c_0 \sum_{j \in \mathbb{P}_i} \left[ \left( \rho_j - \rho_i \right) \frac{\mathbf{r}_{ij} \cdot \nabla_i W_{ij}}{\vert \mathbf{r}_{ij} \vert^2} \frac{m_j}{\rho_j} \right]\,.
\label{eq:sphrho}
\end{equation}

The dimensionless coefficient $0 < \delta \leqslant 1$ adjusts the intensity of the artificial diffusive term. In this work, $\delta = 0.1$ was adopted for all simulations with WCSPH.

Following the work of \citeauthor{monaghan94} \shortcite{monaghan94}, the flow is considered as weakly-compressible and an equation of state is used to determine fluid pressure based on the particle density. The Tait equation of state is commonly used:

\begin{equation}
P_i = \frac{\rho_0 c_0^2}{\gamma} \left[ \left( \frac{\rho_i}{\rho_0} \right)^\gamma - 1 \right]\,.
\label{eq:tait}
\end{equation}

In the above equation, the polytrophic index $\gamma = 7$ is typically adopted for fluid phase, $\rho_0$ is the reference density and $c_0$ is an artificial speed of sound. The artificial speed of sound is adopted instead of the real speed one $c = \sqrt{\frac{\partial P}{\partial \rho}}$ because the later require a very small time step. In order to keep density variations less than $1\%$, the value of $c_0$ is chosen about ten times the maximum velocity $\vert\mathbf{u}\vert_{max}$, according to \citeauthor{monaghan94} \shortcite{monaghan94}.

The discretized governing equations (\ref{eq:sphu}) and (\ref{eq:sphrho}) are integrated using a time-integration scheme based on the Verlet method \cite{verlet67}. At first, all variables are predicted according to:

\begin{equation}
\mathbf{u}_i^{t + \Delta t} = \mathbf{u}_i^{t - \Delta t} + 2 \Delta t \frac{\mathrm{D} \mathbf{u}_i }{\mathrm{D}t}^t \,,
\label{eq:sphu1}
\end{equation}

\begin{equation}
\mathbf{r}_i^{t + \Delta t} = \mathbf{r}_i^t + \Delta t \mathbf{u}_i^t + \frac{\Delta t^2}{2} \frac{\mathrm{D} \mathbf{u}_i }{\mathrm{D}t}^t \,,
\label{eq:sphr1}
\end{equation}

\begin{equation}
\rho_i^{t + \Delta t} = \rho_i^{t - \Delta t} + 2 \Delta t \frac{\mathrm{D} \rho_i }{\mathrm{D}t}^t \,.
\label{eq:sphrho1}
\end{equation}

Aiming to prevent the divergence, i.e., error accumulation, of the values of $\mathbf{u}_i^{t + \Delta t}$, $\mathbf{r}_i^{t + \Delta t}$ and $\rho_i^{t + \Delta t}$ over time, once every $N_s$ time steps (here $N_s = 40$ is adopted), these variables are updated as:

\begin{equation}
\mathbf{u}_i^{t + \Delta t} = \mathbf{u}_i^t + \Delta t \frac{\mathrm{D} \mathbf{u}_i }{\mathrm{D}t}^t \,,
\label{eq:sphu2}
\end{equation}

\begin{equation}
\mathbf{r}_i^{t + \Delta t} = \mathbf{r}_i^t + \Delta t \mathbf{u}_i^t + \frac{\Delta t^2}{2} \frac{\mathrm{D} \mathbf{u}_i }{\mathrm{D}t}^t \,,
\label{eq:sphr2}
\end{equation}

\begin{equation}
\rho_i^{t + \Delta t} = \rho_i^t + \Delta t \frac{\mathrm{D} \rho_i }{\mathrm{D}t}^t \,.
\label{eq:sphrho2}
\end{equation}

In DualSPHysics, the time step of the above equations, here named as $\Delta t_{WCSPH}$, is adaptative and is selected following \citeauthor{monaghan99} \shortcite{monaghan99}, i.e., the minimum of two conditions, the force per unit mass vector $\vert \mathbf{f}_i \vert$, equivalent to the magnitude of particle acceleration (advection), and the viscous term $\Pi_{ij}$ (diffusion):

\begin{equation}
\Delta t_{\text{WCSPH}} = Cr \cdot \min \left\lbrace \min_{\forall i \in \mathbb{P}} \left[ \sqrt{\frac{h}{\vert \mathbf{f}_i \vert}} \right] \,, \:\:\: \min_{\forall i \in \mathbb{P}} \left[ \frac{h}{c_0 + \displaystyle\max_{ j \in \mathbb{P}_i} \left( \frac{h \mathbf{u}_{ij} \cdot \mathbf{r}_{ij}}{\vert \mathbf{r}_{ij} \vert^2 + \eta^2} \right) } \right] \right\rbrace \,,
\label{eq:sphdt}
\end{equation}

where $\mathbb{P}$ denotes all particle domain. The variable $0 < Cr \leq 1$ is the Courant number \cite{courant67}.

Considering that the WCSPH has a fully explicit algorithm, several studies investigating the effects of different time-integration schemes on the accuracy and stability computations have been carried out. The reader interested in such context is invited to refer to \cite{molteni09,mabssout13,violeau14,kolukisa17}.

The dynamic boundary conditions described in \citeauthor{crespo07} \shortcite{crespo07} are used in this work. The boundary particles satisfy the continuity equation (\ref{eq:mass}) as the fluid particles, therefore, their density and pressure are also computed, although they do not move according to the forces exerted on them. Hence, when a fluid particle approaches the boundary particles, and get inside their kernel range, the density of the later increases, leading to a pressure increase. Consequently, the force exerted on the fluid particle increases due to the pressure term in the momentum equation (\ref{eq:momentum}) creating a repulsive force between fluid and boundary particles.

For a more detailed description regarding the characteristics of the open source DualSPHysics, we forward the reader to \citeauthor{crespo15} \shortcite{crespo15}.

\subsection{Moving particle semi-implicit (MPS)}
In the MPS method, the differential operators of the governing equations of continuum are replaced by discrete operators derived based on a weight function $\omega_{ij}$, which accounts the influence of a particle $j$ in the neighborhood of the $i$-th particle, and is given by:

\begin{equation}
\omega_{ij} = \left\{
\begin{array}{cl}
\frac{r_e}{\vert \mathbf{r}_{ij} \vert} - 1 & \:\:\: \vert \mathbf{r}_{ij} \vert \leq r_e \\
0 & \:\:\: \vert \mathbf{r}_{ij} \vert > 0
\end{array} \right. \,,
\label{eq:mpsw}
\end{equation}

where $r_e = k_{MPS} l^0$ is the effective radius that limits the range of influence, $k_{MPS}$ is a constant ranging in [2,4] \cite{koshizuka96} and $ \vert \mathbf{r}_{ij} \vert = \vert \mathbf{r}_j - \mathbf{r}_i \vert $ is the distance between $i$ and $j$.

The summation of the weight of all the particles in the neighborhood $\mathbb{P}_i$ of the $i$-th particle is defined as its particle number density $n_i$, which is proportional to the fluid density:

\begin{equation}
n_i = \sum_{j \in \mathbb{P}_i} \omega_{ij}\,.
\label{eq:mpspnd}
\end{equation}

For an arbitrary function $\phi$, the gradient and Laplacian operators are, respectively, defined as:

\begin{equation}
\langle \nabla \phi \rangle_i = \frac{dim}{n^0} \sum_{j \in \mathbb{P}_i} \left( \frac{\phi_j - \phi_i}{\vert \mathbf{r}_{ij} \vert^2} \mathbf{r}_{ij} \omega_{ij} \right)\,,
\label{eq:mpsgrad}
\end{equation}

\begin{equation}
\langle \nabla^2 \phi \rangle_i = \frac{2dim}{\lambda^0 n^0} \sum_{j \in \mathbb{P}_i} \left[ \left(\phi_j - \phi_i \right) \omega_{ij} \right]\,,
\label{eq:mpslap}
\end{equation}

where $dim$ denotes the number of spatial dimensions and $n^0$ represents the constant particle number density for a fully filled compact support. Finally, $\lambda^0$ is a correction parameter so that the variance increase is equal to that of the analytical solution, and is calculated by:

\begin{equation}
\lambda^0 = \frac{\sum_{j \in \mathbb{P}_i} \left( \vert \mathbf{r}_{ij}^0 \vert^2 \omega_{ij}^0 \right)}{\sum_{j \in \mathbb{P}_i} \omega_{ij}^0}\,.
\label{eq:mpslambda}
\end{equation}

To solve the incompressible viscous flow, a semi-implicit algorithm is used in the MPS method. At first, predictions of the particle’s velocity ($\mathbf{u}_i^*$) and position ($\mathbf{r}_i^*$) are carried out explicitly by using viscosity and external forces terms of the momentum conservation, see Eq.~(\ref{eq:momentum}):

\begin{equation}
\mathbf{u}_i^* = \mathbf{u}_i^t + \Delta t \left( \nu \langle \nabla^2 \mathbf{u} \rangle_i + \mathbf{f}_i \right)^t \,,
\label{eq:mpsu1}
\end{equation}

\begin{equation}
\mathbf{r}_i^* = \mathbf{r}_i^t + \Delta t \mathbf{u}_i^* \,.
\label{eq:mpsr1}
\end{equation}

Then the pressure of all particles is calculated by the PPE as follows \cite{koshizuka99,ikeda01}:

\begin{equation}
\langle \nabla^2 P \rangle_i^{t + \Delta t} - \frac{\rho}{\Delta t^2} \kappa_c P_i^{t + \Delta t} = -\kappa_r \frac{\rho}{\Delta t^2} \frac{n_i^* - n^0}{n^0} \,,
\label{eq:mpslapp}
\end{equation}

where $n_i^*$ stands the particle number density calculated based on the displacement of particles obtained in the prediction step, $\kappa_c$ denotes the coefficient of artificial compressibility and $\kappa_r$ is the relaxation coefficient. In this work, the resulting linear system of the PPE is solved by the conjugate gradient (CG) method with a relative tolerance of $1.0 \times 10^{-6}$. Both $\kappa_c$ and $\kappa_r$ are used to improve the stability of the computation method. The term involving the coefficient $\kappa_c$ leads to a diagonal dominant matrix accelerating the convergence for solving the linear system of the PPE, but also allows a limited compression. Generally, a very small coefficient $\kappa_c \approx \mathcal{O}(10^{-8})$ is adopted so that the compressibility is strictly restricted, which is followed in the present study.

In order to prevent instability issues induced by attractive pressure and reduces the effect of nonuniform particle distribution, we adopted the first order pressure gradient \cite{wang17}:

\begin{equation}
\langle \nabla P \rangle_i = \left[ \sum_{j \in \mathbb{P}_i} \left( \frac{\mathbf{r}_{ij}}{\vert \mathbf{r}_{ij} \vert} \otimes \frac{\mathbf{r}_{ij}}{\vert \mathbf{r}_{ij} \vert} \omega_{ij} \right) \right]^{-1} \sum_{j \in \mathbb{P}_i} \left( \frac{P_j - \widehat{P}_i}{\vert \mathbf{r}_{ij} \vert^2} \mathbf{r}_{ij} \omega_{ij} \right) \,,
\label{eq:mpsgradp}
\end{equation}

where $\widehat{P}_i$ is the minimum pressure between the neighborhood $\mathbb{P}_i$ of the particle $i$.

Finally, the velocity of the particles is updated by:

\begin{equation}
\mathbf{u}_i^{t + \Delta t} = \mathbf{u}_i^* - \frac{\Delta t}{\rho} \langle \nabla P \rangle_i^{t + \Delta t} \,,
\label{eq:mpsu2}
\end{equation}

and the new positions of the particles are corrected by:

\begin{equation}
\mathbf{r}_i^{t + \Delta t} = \mathbf{r}_i^* + \Delta t \left( \mathbf{u}_i^{t + \Delta t} - \mathbf{u}_i^* \right) \,.
\label{eq:mpsr2}
\end{equation}

For the MPS simulations, a fixed time step, named as $\Delta t_{MPS}$, and based on the maximum velocity $\vert \mathbf{u} \vert_{max}$, is initially assigned following the Courant-Friedrichs-Lewy (CFL) condition:

\begin{equation}
\Delta t_{MPS} \leq \frac{Cr \:\: l^0}{\vert \mathbf{u} \vert_{max}}\,.
\label{eq:mpsdt}
\end{equation}

In the present work, the Euler implicit method is adopted as time-integration scheme on the MPS. Although the Euler implicit method is a first-order time-integration scheme, since a predictor–corrector scheme is used to solve the governing equations, its adoption is accurate enough for the MPS simulations, as pointed out by \citeauthor{shimizu16} \shortcite{shimizu16}.

\begin{figure}[th]
\centerline{\includegraphics[width=10cm]{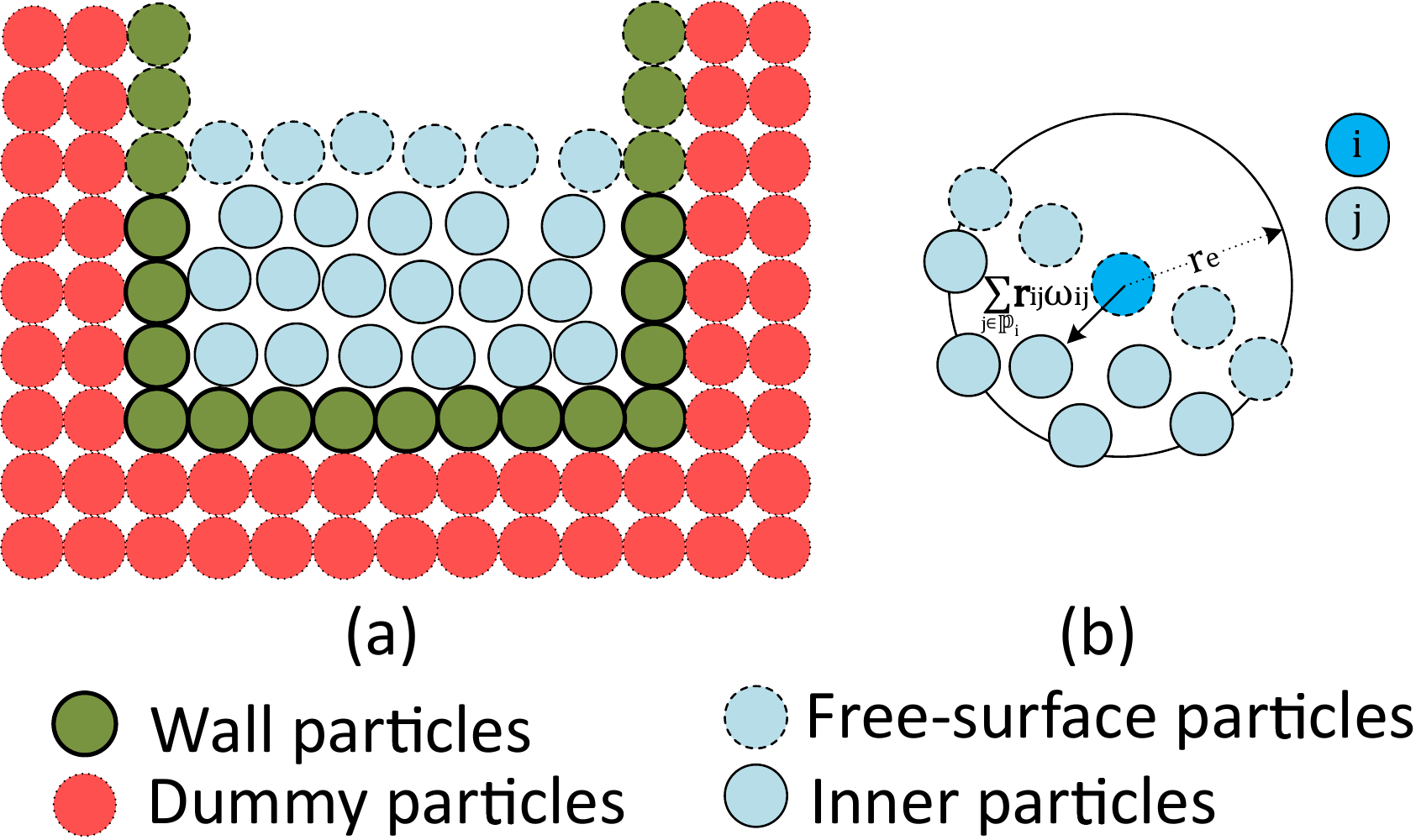}}
\vspace*{8pt}
\caption{Boundary conditions in 2D space. (a) Fluid particles and solid wall modeled by wall and dummy particles. (b) Free-surface particles: resultant weighted deviation vector.\label{one}}
\vspace*{-\baselineskip}
\end{figure}

Solid wall is modeled by using three layers of fixed particles, as shown in Fig.~\ref{one}(a). The particles that form the layer in contact with the fluid are denominated wall particles, of which the pressure is computed by solving the PPE, see Eq.~(\ref{eq:mpslapp}), together with the fluid particles. The particles that form two other layers are denominated dummy particles, which are used to ensure the correct calculation of the particle number density of the wall particles. Pressure is not calculated in the dummy particles. As boundary condition of rigid walls, the no-slip condition is considered on the wall. The Dirichlet pressure boundary condition is imposed to the free-surface particles and it is considered during the implicit step of the method. The neighborhood particles centroid deviation (NPCD) method \cite{tsukamoto16} is adopted in the present work. By accurately identifying free-surface particles, the NPCD method improves the stability and accuracy of the pressure computation, and eliminates spurious oscillations due to misdetection of free-surface particles inside the fluid domain. It provides satisfactory results even by using only the particle number density deviation as the source term in Eq.~(\ref{eq:mpslapp}), as verified in \citeauthor{tsukamoto16} [\citeyear{tsukamoto16}; \citeyear{tsukamoto20}]. NCPD is a two criteria method to detect free-surface particles, consisting in the following procedure: 

\begin{arabiclist}
\item particle number density criterion originally proposed by \citeauthor{koshizuka96} \shortcite{koshizuka96}, which check the fullness of the neighborhood:

\begin{equation}
n_i < \beta_1 \: n^0\,,
\label{eq:mpsni}
\end{equation}

\item deviation of the centroid of the neighborhood particles, which check the symmetry of the particle distribution in the neighborhood:

\begin{equation}
\sigma_i > \varrho_1 \: l^0\,,
\label{eq:mpsisgmai1}
\end{equation}

\end{arabiclist}

with the deviation $\sigma_i$ calculated as:

\begin{equation}
\sigma_i = \frac{ \sqrt{ \left[ \sum_{j \in \mathbb{P}_i} \left( x_{ij} \: \omega_{ij} \right) \right]^2 + \left[ \sum_{j \in \mathbb{P}_i} \left( y_{ij} \: \omega_{ij} \right) \right]^2 + \left[ \sum_{j \in \mathbb{P}_i} \left( z_{ij} \: \omega_{ij} \right) \right]^2}}{\sum_{j \in \mathbb{P}_i} \omega_{ij} } \,,
\label{eq:mpssigmai2}
\end{equation}

where $x_{ij} = (x_j-x_i)$, $y_{ij} = (y_j-y_i)$ and $z_{ij} = (z_j-z_i)$. If the latter two criteria are satisfied, the particle is identified as free-surface one and its pressure is set to zero.

The constant $\beta_1$ should be chosen between 0.80 and 0.99 \cite{koshizuka96} and $\varrho_1$ higher than 0.2 \cite{tsukamoto16}. The values of $\beta_1 = 0.98$ and $\varrho_1 = 0.2$ are used for all simulations with MPS performed herein. Figure~\ref{one}(b) shows the main idea behind the NPCD method, i.e., the resultant weighted deviation vector.

\section{Results}
Two 3D dam-break cases are studied. First, computed wave heights and pressures are compared to experimental \cite{kleefsman05} and numerical ones \cite{vaz09}. After that, computed and experimental \cite{aureli15} forces on a box are compared for a second dam-break case.
For all simulations, we adopted the density of the fluid as $\rho = 1000 kg/m^3$, and the gravity acceleration $g = 9.81 m/s^2$. The artificial speed of sound $c_0=20\sqrt{gH_W}$, where $H_W$ is the initial water column height, was used in WCSPH. The coefficients $\kappa_c = 10^{-8} ms^2/kg$ and $\kappa_r = 0.01$ were found appropriate based on some trial tests and thus being adopted for the MPS, see Eq.~(\ref{eq:mpslapp}). Regarding the time step, the Courant number $Cr \approx 0.2$ usually ensures the stability of the calculation of trial numerical experiments. Therefore, $Cr = 0.2$ was assigned for both methods.

\subsection{3D dam-break flow - Wave height and pressure}

\begin{figure}[th]
\centerline{\includegraphics[width=13cm]{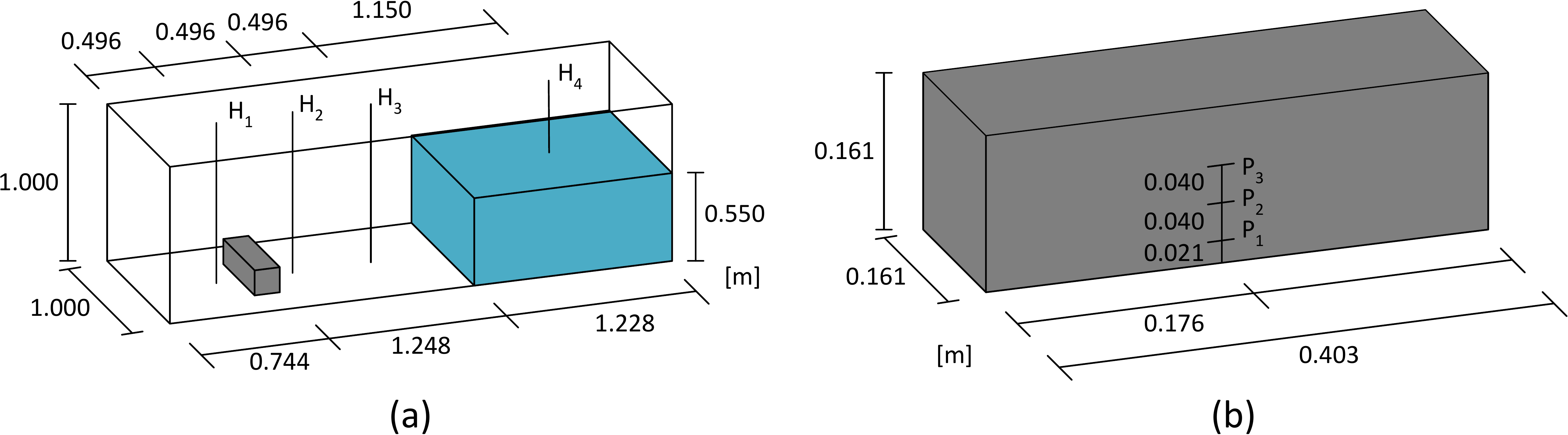}}
\vspace*{8pt}
\caption{(a) Initial geometry, main dimensions, wave probes (H$_1$, H$_2$, H$_3$ and H$_4$) and (b) pressure sensors (P$_1$, P$_2$ and P$_3$) on the box \protect\cite{kleefsman05}.\label{two}}
\vspace*{-\baselineskip}
\end{figure}

The first case (Case 1) is based on the experiment performed by \citeauthor{kleefsman05} \shortcite{kleefsman05}. The initial geometry of the tank of length $L_T = 3.22 m$, width $W_T = 1.00 m$ and height $H_T = 1.00 m$, box of length $L_B = 0.161 m$, width $W_B = 0.403 m$ and height $H_B = 0.161 m$ and water column of initial height $H_W = 0.55 m$, as well as the wave probes positions (H$_1$, H$_2$, H$_3$ and H$_4$) and pressure sensors (P$_1$, P$_2$ and P$_3$) on the box, are shown in Fig.~\ref{two}.

Similar to the investigation performed for a 2D dam-break flow in \citeauthor{bakti16} \shortcite{bakti16}, four different artificial viscosity values were adopted here for the 3D simulations with WCSPH, and their effects on the free-surface flow and pressure on the box are discussed. The artificial viscosity is generally adopted to smooth the unphysical numerical oscillations, although it can also introduce undesirable numerical problems, e.g., excess dissipation and false shearing torque in rotating flows \cite{flebbe94}. According to \citeauthor{monaghan05} \shortcite{monaghan05}, the continuum limit of the viscosity shows that, for the Wendland kernel, see Eq.~(\ref{eq:sphw}), the kinematic ($\nu$) and artificial ($\alpha$) viscosities can be related by:

\begin{equation}
\nu_i = \frac{\alpha_i h c_i}{2 \left(dim + 2\right)}\,.
\label{eq:visci}
\end{equation}

Assuming the artificial speed of sound almost constant $c_0$, as done in \citeauthor{bakti16} \shortcite{bakti16}, the relation

\begin{equation}
\nu_0 = \frac{\alpha_0 h c_0}{2 \left(dim + 2\right)}
\label{eq:visc0}
\end{equation}

was used here to correlate kinematic viscosity and the artificial viscosity adopted in WCSPH. Numerical parameters and processing times are presented in Table~\ref{tab1}.

\begin{table}[th]
\tbl{Numerical parameters and processing time of Case 1.\label{tab1}}
{\begin{tabular}{@{}cccc@{}} \toprule
Parameter & WCSPH & MPS \\ \colrule
\multirow{2}{6em}{Viscosity ($\nu_0$)} & $5.2 \times 10^{-3}$, $5.2 \times 10^{-4}$, & \multirow{2}{4em}{$10^{-6} \: m^2/s$} \\
 & $5.2 \times 10^{-5}$, $5.2 \times 10^{-6} \: m^2/s$ & \\
Artificial viscosity ($\alpha_0$) & 0.1, 0.01, 0.001, 0.0001 & - \\
Particle distance ($\l^0$) & $0.0075 \: m$ & $0.0075 \: m$ \\
Radius of support & $2h = 3l^0$ & $r_e = 2.1l^0$ \\
Fluid particles & $1.6 \times 10^6$ & $1.6 \times 10^6$ \\
Solid particles & $2.1 \times 10^5$ & $6.5 \times 10^5$ \\
Simulation time & $6 \: s$ & $6  \: s$ \\
Time step ($\Delta t$) & $3.8 \times 10^{-5} \: s^{a}$ & $3.0 \times 10^{-4}  \: s$ \\
Computational time & (CPU 1d11h00m$^b$)(GPU 0d4h00m$^c$) & (CPU 2d14h00m$^b$) \\ \botrule
\end{tabular} }
\begin{tabfootnote}
\tabmark{a} Based on the number of computational steps. DualSPHysics uses an adaptative time step restrict by advection and diffusion stability criteria (see Eq.~(\ref{eq:sphdt}))\\
\tabmark{b} CPU Intel \textsuperscript{\textregistered} Xeon \textsuperscript{\textregistered} Processor E5 v2 Family, processor base frequency of 2.80 GHz, 20 cores and 126 GB of memory\\
\tabmark{c} GPU Nvidia \textsuperscript{\textregistered} Tesla \textsuperscript{\textregistered} K40m, 15 Multiprocessors (2880 cores), clock rate of 0.75 GHz and 11.519 GB of global memory
\end{tabfootnote}
\end{table}

\begin{figure}[th]
\centerline{\includegraphics[width=13cm]{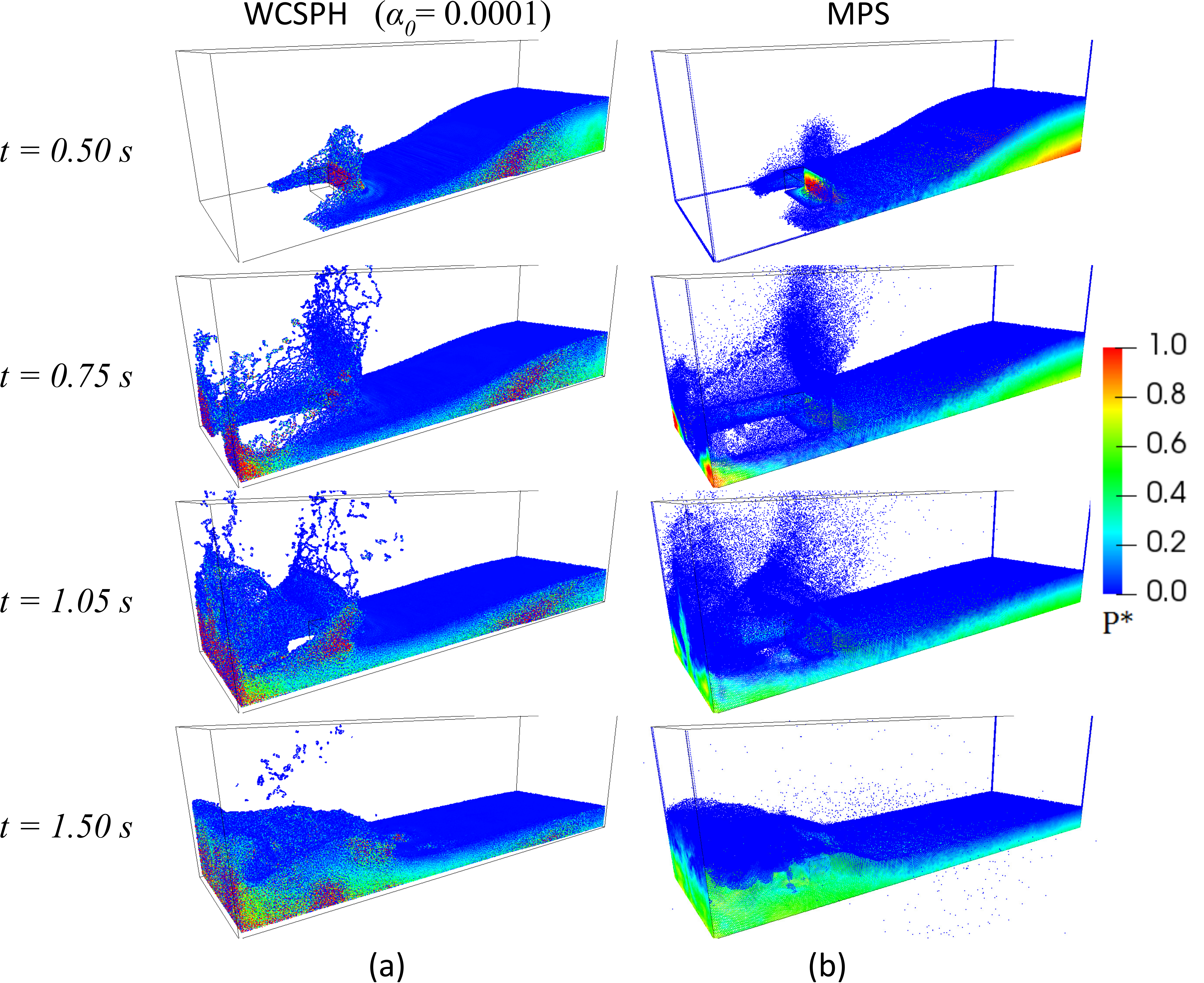}}
\vspace*{8pt}
\caption{Snapshots of the dimensionless pressure distribution of dam-break simulations carried out with (a) WCSPH adopting $\alpha_0 = 0.0001$ and (b) MPS.\label{three}}
\vspace*{-\baselineskip}
\end{figure}

The snapshots of the free-surface deformation computed by the WCSPH ($\alpha_0 = 0.0001$) and MPS are illustrated in Fig.~\ref{three}. Only free-surface particles and wall particles with positive pressure are shown in the simulations. The colors on the free-surface and wall particles are related to its dimensionless pressure magnitude $P^* = \frac{P}{\rho g H_W}$, where $H_W = 0.55 m$ is the initial water column height. After hitting the weather side of the box, large splash is formed at the instant $t = 0.50 s$, and part of the wave wraps the box. The wave collides the corners of the tank wall, and the splash reaches the top of the domain at the instant $t = 0.75 s$, approximately. After that, at the instant $t = 1.05 s$, the reflected wave breaks and the splashed fluid fall due the gravity, followed by their merging near the box at the instant $t = 1.50 s$, approximately. Overall, both methods produce a similar fluid behavior, despite the lower spreading of the splash computed by the WCSPH at the instant $t = 1.05 s$. This difference might be attributed to the artificial viscosity in WCSPH, although a relatively low value was adopted here. On the other hand, one can argue that in fact the particles in MPS are more scattered than in the WCSPH, since this scattering has been noted in previous results from incompressible particle-based methods \cite{rafiee12}. Concerning the pressure field, the MPS simulation has reduced pressure fluctuations, thereby providing a more realistic pressure distribution than that obtained by applying WCSPH. Furthermore, a much smoother pressure distribution in the fluid is computed by MPS, as pointed by the well represented pressure layers.

\begin{figure}[th]
\centerline{\includegraphics[width=8.5cm]{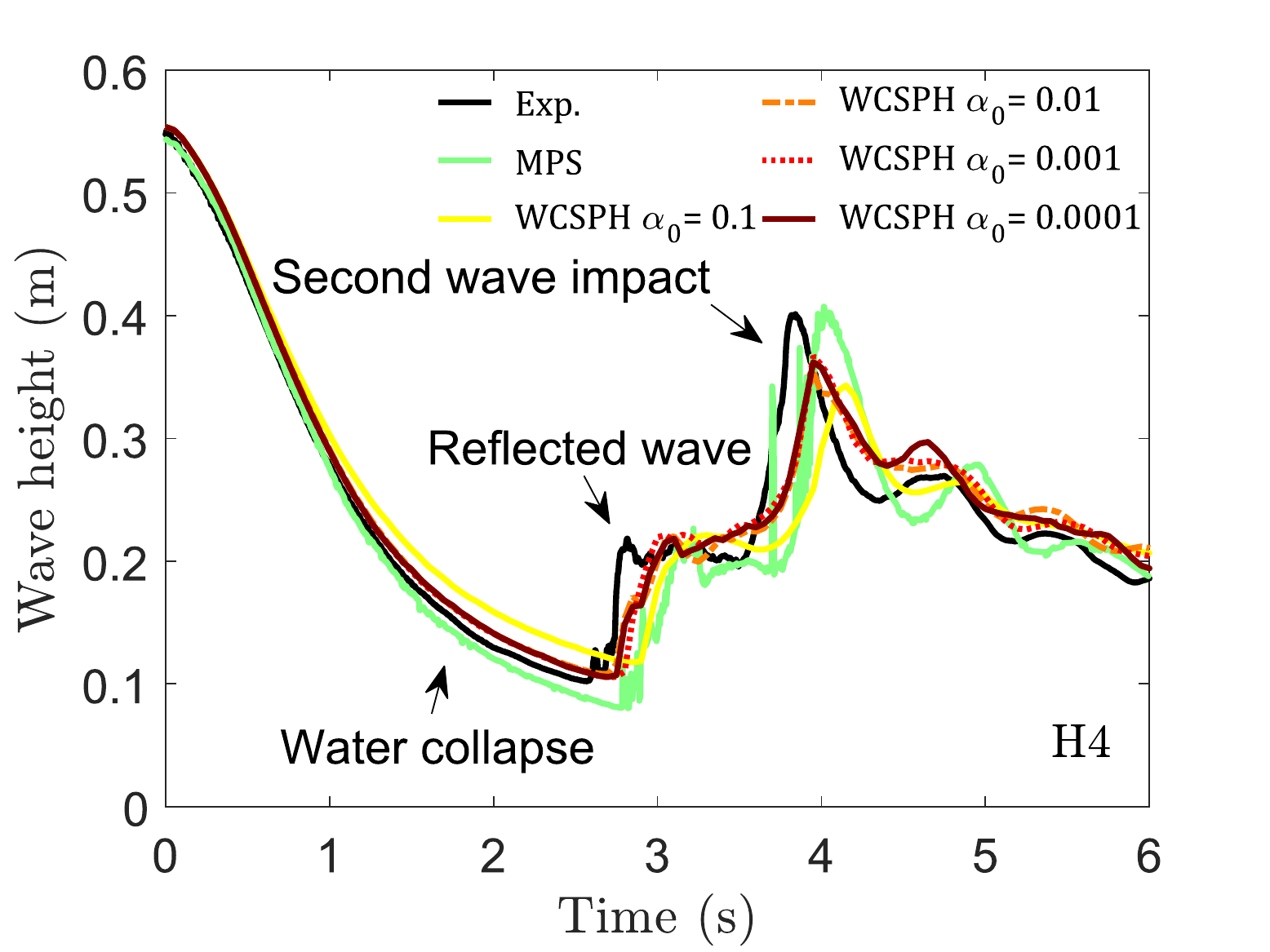}}
\vspace*{8pt}
\caption{Wave height at probe H$_4$. Experimental data \protect\cite{kleefsman05} and numerically computed using MPS and WCSPH.\label{four}}
\vspace*{-\baselineskip}
\end{figure}

Figure~\ref{four} shows the wave heights at the probe H$_4$. Compared to the experimental data, the water column collapsing occurs slightly slower for the WCSPH simulations with $\alpha_0 \leq 0.01$ and slightly faster for the MPS simulation. For the WCSPH simulation with $\alpha_0 = 0.1$, the delay is remarkable. All computed heights present a delay after the first reflected wave, at approximately $t = 2.7 s$. As reported in \citeauthor{moulinec08} \shortcite{moulinec08}, these deviations might be related to the interaction with the boundary during the impact on the back and front walls, so that WCSPH might overestimate the boundary effect on the flow. Moreover, since the conventional WCSPH, which is used in the present work, presents low accuracy for wave breaking problems with highly irregular particle distributions, it is reasonable to suppose that improved versions, such as WCSPH based on a low-dissipation Riemann solver \cite{zhang17}, the kernel gradient correction (KGC) SPH (\citeauthor{shao12}, \citeyear{shao12}; \citeyear{shao16}) or the kernel gradient-free (KGF) SPH \cite{huang19} may decrease these deviations given that higher accuracy is expected. On the other hand, the deviations between experimental and numerical wave events occur in the MPS simulation due to an inherent numerical dissipation. In addition to the underestimated wave height computed by WCSPH during the second incoming wave, the numerical results are delayed in comparison with the experimental one, especially for WCSPH with $\alpha_0 = 0.1$, approximately at the instant $t = 4.5 s$.

\begin{figure}[th]
\centerline{\includegraphics[width=8.5cm]{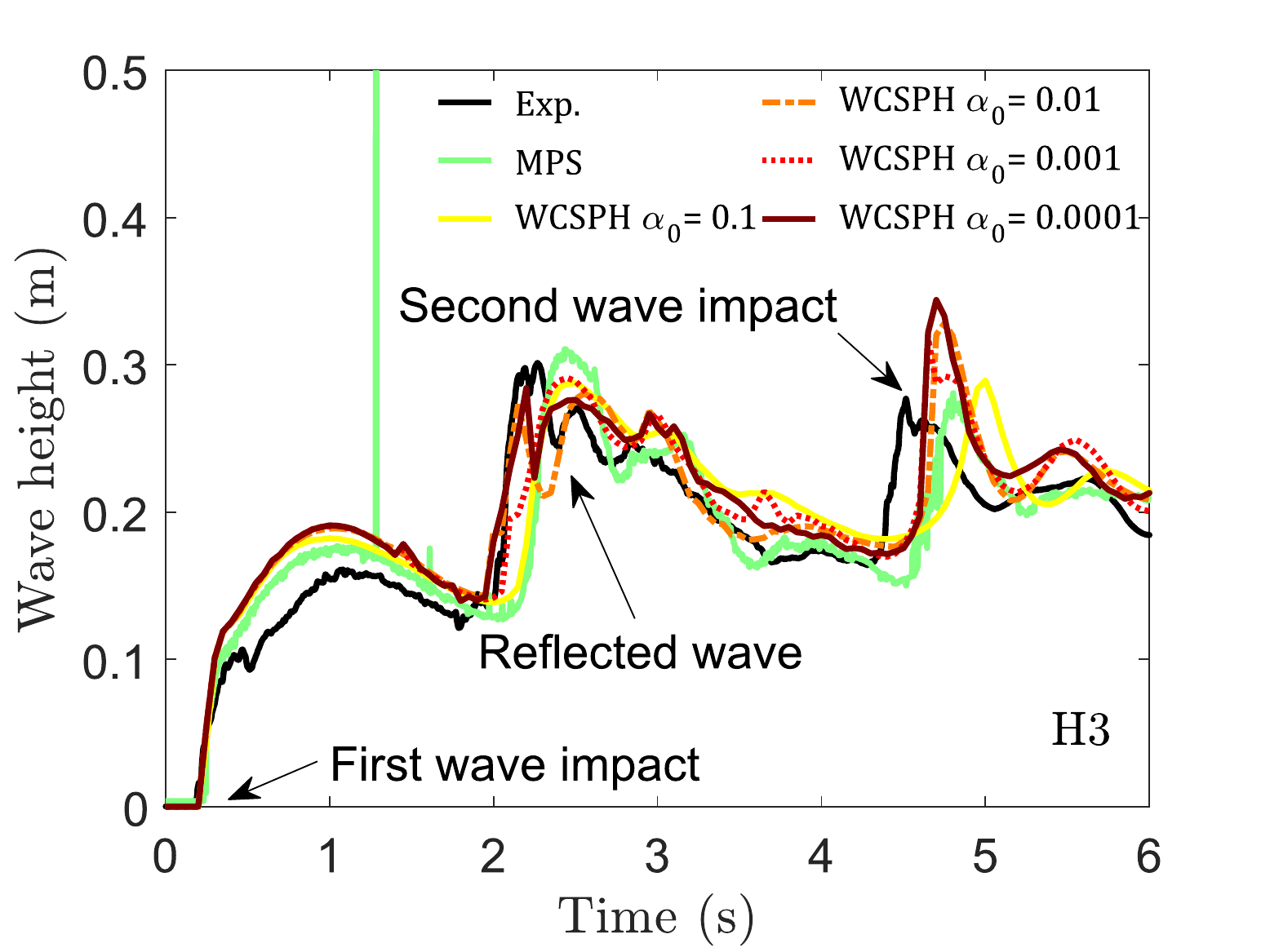}}
\vspace*{8pt}
\caption{Wave height at probe H$_3$. Experimental data \protect\cite{kleefsman05} and numerically computed using MPS and WCSPH.\label{five}}
\vspace*{-\baselineskip}
\end{figure}

Figure~\ref{five} illustrates the wave heights at the probe H$_3$. Until the instant $t = 2.0 s$, the computed wave heights are slightly higher than the experimental one. After that, the wave heights for both methods are in reasonable agreement with the experimental data. However, exceptions occur around the second wave impact, $t = 4.8 s$, where the WCSPH with $\alpha_0 \leq 0.01$ leads to higher waves and, as previous observed, there is a delay in relation to the experimental data.

\begin{figure}[th]
\centerline{\includegraphics[width=8.5cm]{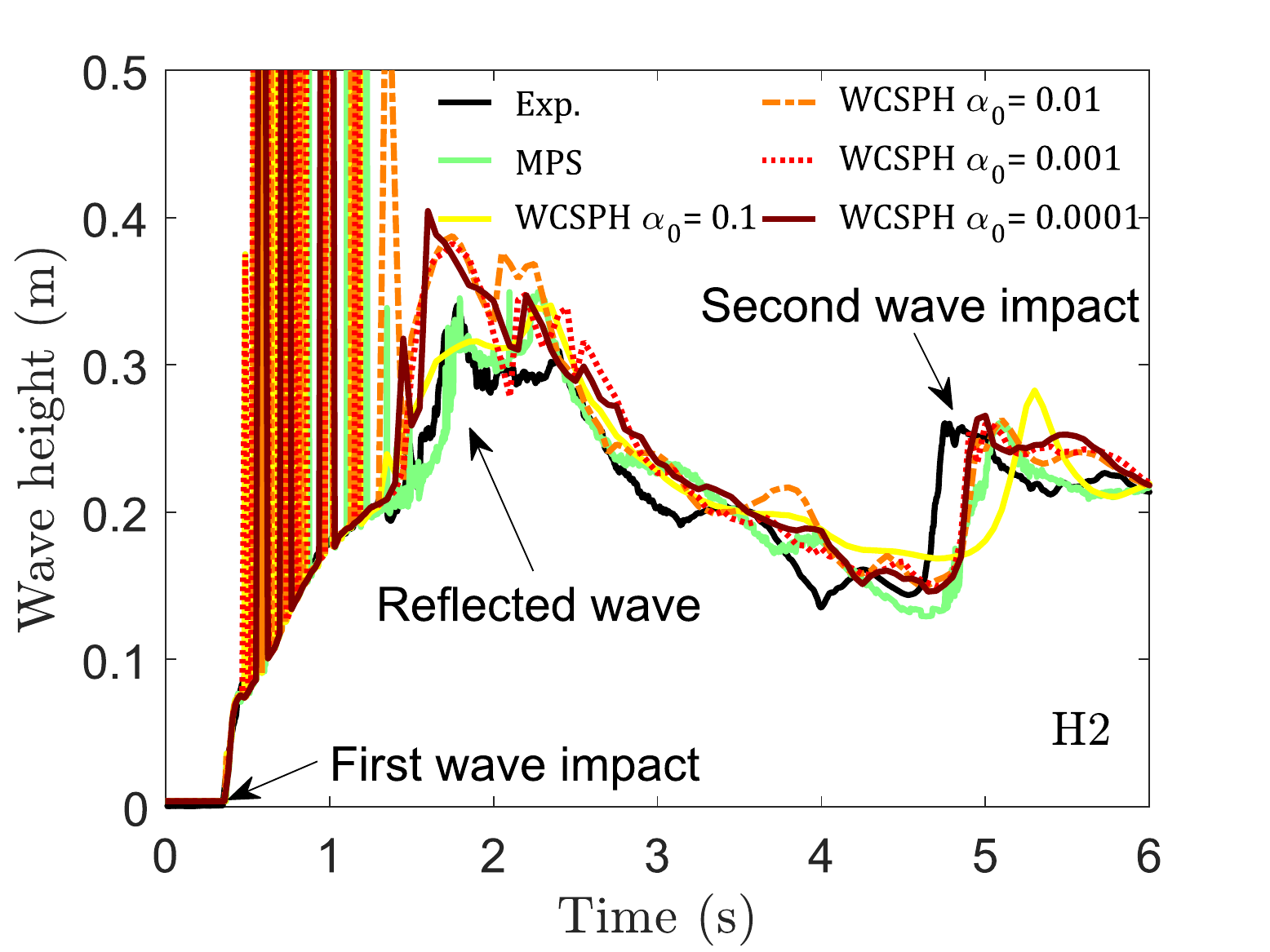}}
\vspace*{8pt}
\caption{Wave height at probe H$_2$. Experimental data \protect\cite{kleefsman05} and numerically computed using MPS and WCSPH.\label{six}}
\vspace*{-\baselineskip}
\end{figure}

The wave heights at the probe H$_2$ are provided in Fig.~\ref{six}. The high values of the computed wave heights from $t = 0.75 s$ to $t = 1.05 s$ reflect the splashing fluid particles that reach the top of the domain. After the wave reflection, between the instants $t = 1.5 s$ and $t = 2.5 s$, the wave heights computed by WCSPH with $\alpha_0 \leq 0.01$ are slighter higher than the experimentally measured one. As previous discussed for the probe H$_3$, the computed wave heights are in reasonable agreement with the experimental data, with exception of the instants after $t = 4.8 s$.

Overall, experimentally measured and numerically computed wave heights are in good agreements, except for WCSPH with the artificial viscosity $\alpha_0 \geq 0.1$, or $\nu_0 \geq \mathcal{O}(10^{-2})$.

\begin{figure}[th]
\centerline{\includegraphics[width=8.5cm]{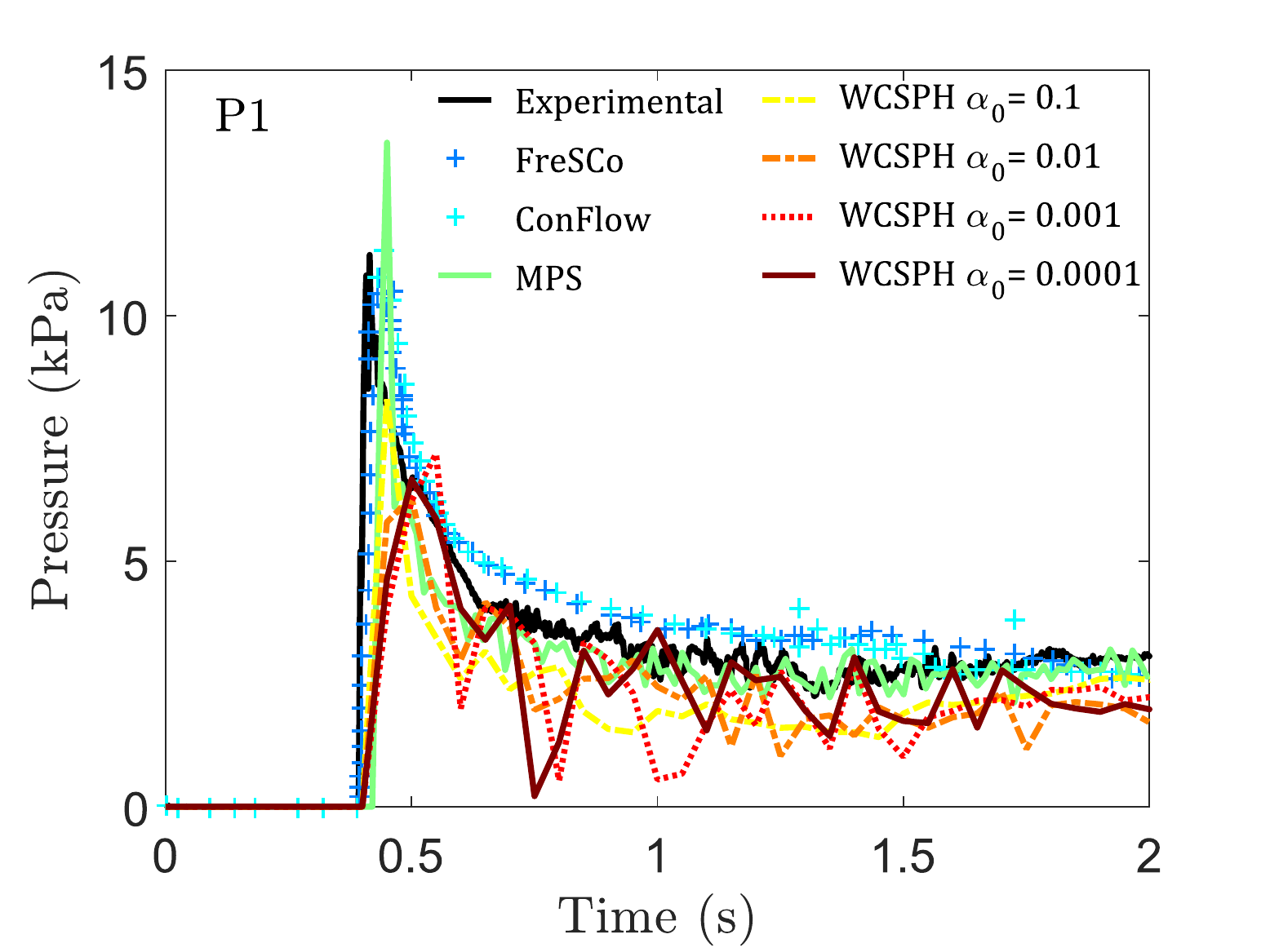}}
\vspace*{8pt}
\caption{Pressure time series at sensor P$_1$. Experimental data \protect\cite{kleefsman05} and numerically computed using FresCo \protect\cite{vaz09}, ConFlow \protect\cite{kleefsman05}, MPS and WCSPH.\label{seven}}
\vspace*{-\baselineskip}
\end{figure}

\begin{figure}[th]
\centerline{\includegraphics[width=8.5cm]{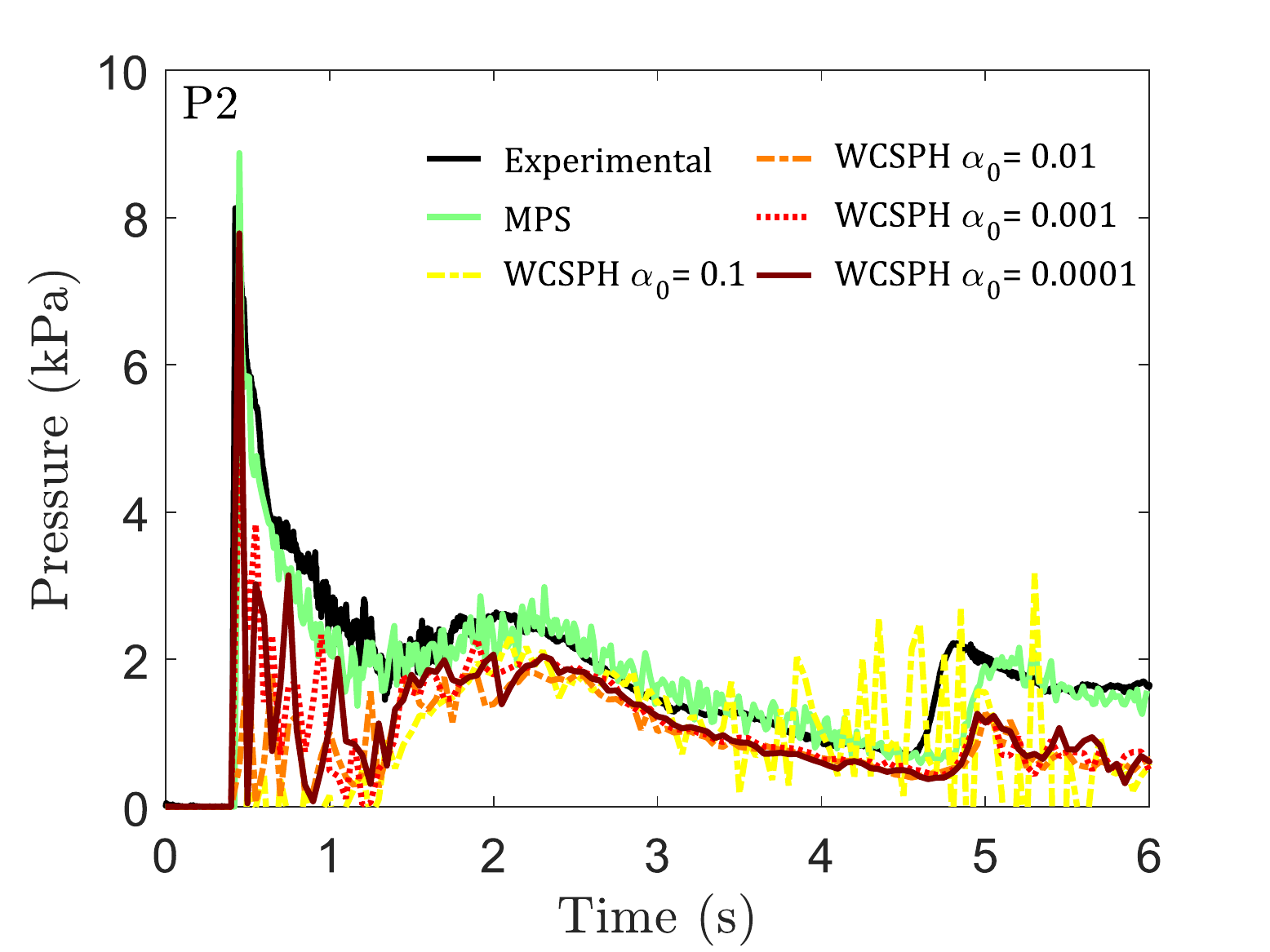}}
\vspace*{8pt}
\caption{Pressure time series at sensor P$_2$. Experimental data \protect\cite{kleefsman05} and numerically computed using FresCo \protect\cite{vaz09}, ConFlow \protect\cite{kleefsman05}, MPS and WCSPH.\label{eight}}
\vspace*{-\baselineskip}
\end{figure}

\begin{figure}[th]
\centerline{\includegraphics[width=8.5cm]{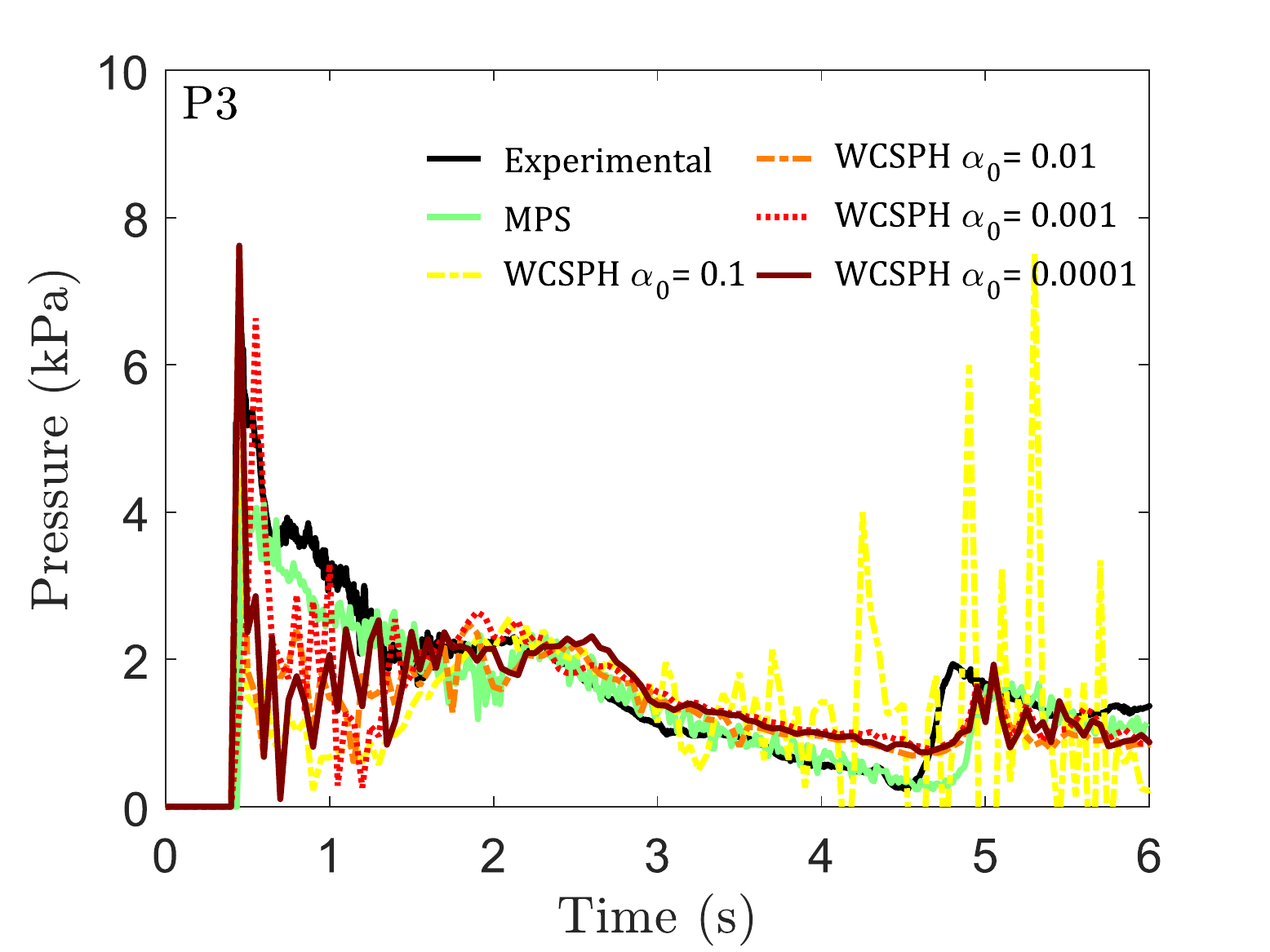}}
\vspace*{8pt}
\caption{Pressure time series at sensor P$_3$. Experimental data \protect\cite{kleefsman05} and numerically computed using FresCo \protect\cite{vaz09}, ConFlow \protect\cite{kleefsman05}, MPS and WCSPH.\label{nine}}
\vspace*{-\baselineskip}
\end{figure}

In Figs.~\ref{seven}, \ref{eight} and \ref{nine}, the computed pressure time series at P$_1$, P$_2$ and P$_3$, are compared with the experimental data \cite{kleefsman05}, respectively. In addition, numerical results by the mesh-based codes FreSCo \cite{vaz09} and ConFlow \cite{kleefsman05}, both adopting volume of fluid (VOF) approach to handle free-surfaces, are shown. First, it is important to clarify how the computed pressure time series are obtained in each particle-based method. In the present MPS, raw data of the pressure calculated at the wall particle that is closest one to the sensor position is considered. In WCSPH, the default output pressure of DualSPHysics is given by:

\begin{equation}
P_a = \frac{\sum_{j \in \mathbb{P}_a} \left( P_b \: W_{ab} \right)}{\sum_{j \in \mathbb{P}_a} W_{ab}}\,,
\label{eq:sphpa}
\end{equation}

where $P_b$ is the pressure of fluid particles $b$ in the neighborhood of the point $a$. Although Eq.~(\ref{eq:sphpa}) gives the spatial average pressure, instead the raw data at a specific point, it provides acceptable results for practical purposes.

The peak pressure at sensor P$_1$ is over predicted by MPS while is under predicted by all simulations with WCSPH. On the other hand, computed peak pressure at sensor P$_2$ is in very well agreement with experimental data for MPS and WCSPH with $\alpha_0 = 0.0001$. For the sensor P$_3$, the peak pressure is well predicted by all WCSPH simulations but under predicted by MPS. One aspect to highlight here is that the computed impact pressure peak is highly sensitive to numerical parameters. The reason for this is out of the scope of the present work, but it is affected by discrete approximations, e.g., time step and spatial resolution, as well as numerical parameters, such as compressibility, relaxation coefficients, etc. In this way, discrepancies between experimentally measured and numerically computed peak pressures are expected.

After the first pressure peak, the pressures obtained by MPS are in good agreement with experiment ones. On the other hand, high-amplitude pressure oscillations are computed by WCSPH until $t = 1.5 s$. This might be related to the acoustic waves caused by the weakly-compressibility modeling of the WCSPH \cite{gesteira12} and also due the irregular particle distribution, which in turn affects the numerical stability given the low accuracy of the conventional WCSPH \cite{huang19}. After the instant $t = 1.5 s$, the pressures computed by WCSPH with $\alpha_0 \leq 0.01$ are very similar, whereas the pressure oscillations computed by WCSPH with $\alpha_0 = 0.1$ increase after $t = 3.5 s$. In this way, $\alpha_0 \leq 0.01$, i.e., $\nu_0 \leq \mathcal{O}(10^{-3})$, is preferable for this case.

As the peak values of hydrodynamic impact pressure show some randomness, an alternative reference for the comparison is the use of the impulse. This is because the momentum conservation should be satisfied, independent of the pressure peaks shape. Thus, for sake of simplicity, the pressure impulses $I_p$ computed by WCSPH and MPS during the first 6 seconds were also compared with the experimental ones. The computed pressure impulses and their relative deviations, referred to the experimental ones, can be found in Table~\ref{tab2}. The WCSPH underestimates the pressure impulses with a relative deviation around $29\%$ for the sensor P$_1$. It ranges from $40\%$ to $55\%$ for the sensor P$_2$ and from $8\%$ to $24\%$ for the sensor P$_3$. The increase of the artificial viscosity leads to larger deviations, indicating that $\alpha_0 \leq 0.001$, i.e., $\nu_0 \leq \mathcal{O}(10^{-4})$, is preferable to reproduce the hydrodynamic loads. On the other hand, MPS underestimates the pressure impulses with a relative deviation between $5\%$ and $11\%$. Despite both numerical methods underestimate the experimental results, in terms of hydrodynamic loads, the pressures computed by MPS are more accurate than those obtained by WCSPH.

\begin{table}[th]
\tbl{Pressure impulse $(I_p)$ and deviations at sensors P$_1$, P$_2$ and P$_3$. Relative deviations referred to the experiment.\label{tab2}}
{\begin{tabular}{@{}cccccccccccc@{}} \toprule
Sensor & EXP & \multicolumn{8}{c}{WCSPH} & \multicolumn{2}{c}{MPS} \\ 
 & & \multicolumn{2}{c}{$\alpha_0 = 0.1$} & \multicolumn{2}{c}{$\alpha_0 = 0.01$} & \multicolumn{2}{c}{$\alpha_0 = 0.001$} & \multicolumn{2}{c}{$\alpha_0 = 0.0001$} \\ \colrule
 & $I_p (Pa.s)$ & $I_p (Pa.s)$ & $(\%)$ & $I_p (Pa.s)$ & $(\%)$ & $I_p (Pa.s)$ & $(\%)$ & $I_p (Pa.s)$ & $(\%)$ & $I_p (Pa.s)$ & $(\%)$ \\
P$_1$ & 13392 & 9519 & 29.0 & 9703 & 27.5 & 9988 & 25.4 & 10217 & 23.7 & 12641 & 5.6 \\
P$_2$ & 10961 & 4898 & 55.3 & 5390 & 50.8 & 6416 & 41.5 & 6506  & 40.6 & 10192 & 7.0 \\
P$_3$ & 9500  & 7189 & 24.3 & 7943 & 16.4 & 8791 & 7.5  & 8736  & 8.0  & 8488  & 10.7 \\ \botrule
\end{tabular} }
\end{table}

Concerning the processing CPU time of each method, given in Table~\ref{tab1}, the PPE system solver in MPS play a mandatory role in improving computational efficiency. It is evidenced by the faster computation in the WCSPH simulations, given that $CPUtime_{MPS} \approx 1.8 \times CPUtime_{WCSPH}$. Regarding the CPU and GPU time costs of the WCSPH method, GPU-based calculations is about 8 times faster than CPU-based ones $(CPUtime_{WCSPH} \approx 8 \times GPUtime_{WCSPH})$.

\subsection{3D dam-break flow - Force}

\begin{figure}[th]
\centerline{\includegraphics[width=10cm]{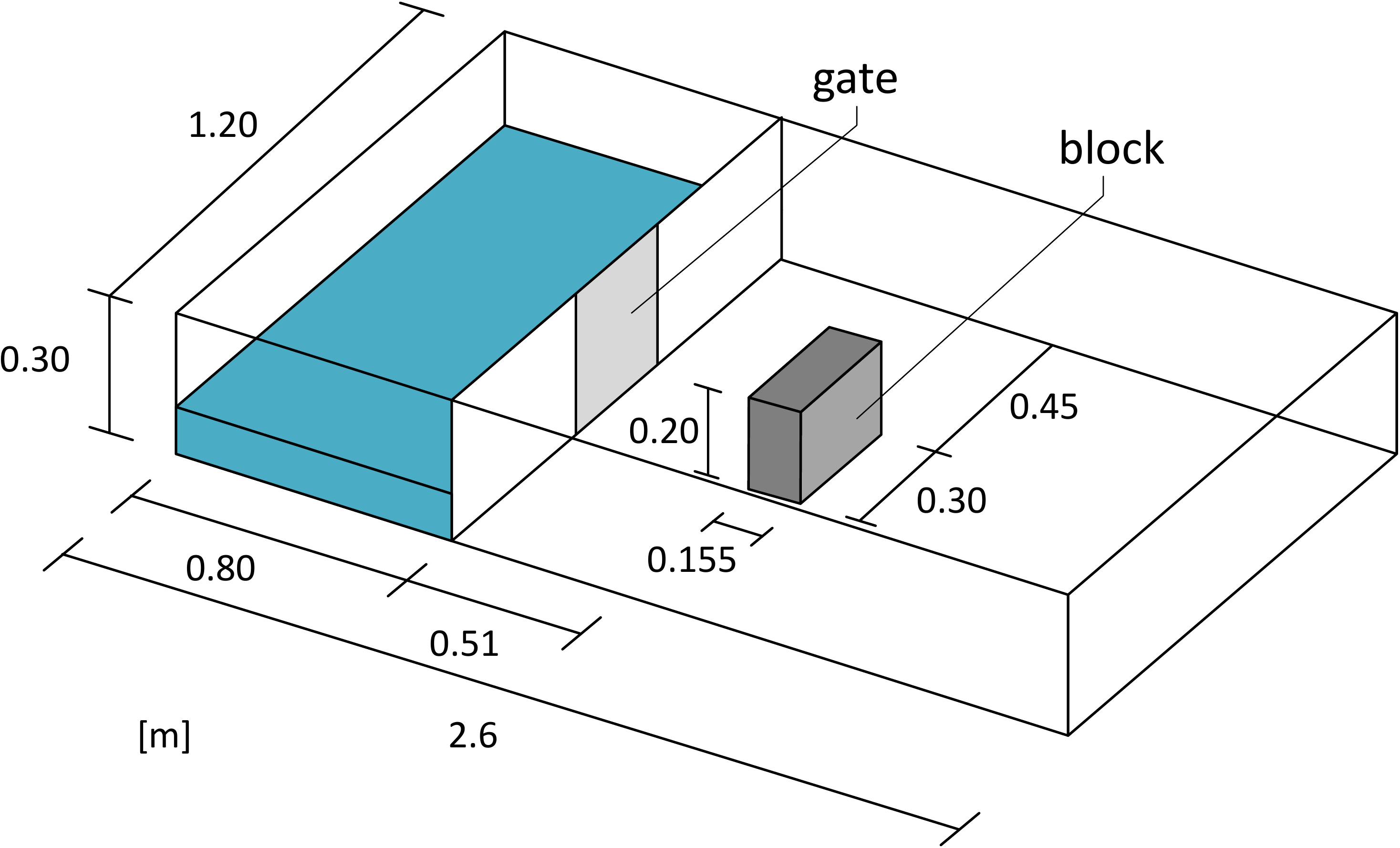}}
\vspace*{8pt}
\caption{Initial geometry and main dimensions \protect\cite{aureli15}.\label{ten}}
\vspace*{-\baselineskip}
\end{figure}

The second case (Case 2) is based on the experiment performed by \citeauthor{aureli15} \shortcite{aureli15}. The initial geometry is shown in Fig.~\ref{ten}. The main dimensions are: tank of length $L_T = 2.60 m$, width $W_T = 1.20 m$ and height $H_T = 0.30 m$, gate of thickness $t_G = 0.025 m$, width $W_T = 0.30 m$ and height $H_G = 0.30 m$ and block of length $L_B = 0.155 m$, width $W_B = 0.30 m$ and height $H_B = 0.20 m$. The water column, of height $H_w = 0.1 m$, is confined by a gate that is opened with a constant vertical velocity of $1.25 m/s$. Two distances between particles, $l^0 = 0.005 m$ and $0.0025 m$, are used for both methods, and three artificial viscosities are adopted for the WCSPH. Numerical parameters and processing times are presented in Table~\ref{tab3}.

\begin{table}[th]
\tbl{Numerical parameters and processing time of Case 2.\label{tab3}}
{\begin{tabular}{@{}ccccc@{}} \toprule
Parameter & \multicolumn{2}{c}{WCSPH} & \multicolumn{2}{c}{MPS} \\ \colrule
\multirow{3}{7em}{Viscosity ($\nu_0$)} & $1.5 \times 10^{-3}$, & $7.4 \times 10^{-3}$, & \multicolumn{2}{c}{$10^{-6} \: m^2/s$} \\
									   & $1.5 \times 10^{-4}$, & $7.4 \times 10^{-4}$, & \\
									   & $1.5 \times 10^{-5} \: m^2/s$ & $7.4 \times 10^{-5} \: m^2/s$ & \\
Artificial viscosity ($\alpha_0$) 	   & \multicolumn{2}{c}{0.1, 0.01, 0.001} & \multicolumn{2}{c}{-} \\
Particle distance ($\l^0$) 			   & $0.005 \: m$ & $0.0025 \: m$ & $0.005 \: m$ & $0.0025 \: m$ \\
Radius of support 					   & \multicolumn{2}{c}{$2h = 3l^0$} & \multicolumn{2}{c}{$r_e = 2.1l^0$} \\
Fluid particles 					   & $7.6 \times 10^5$ & $6.1 \times 10^6$ & $7.7 \times 10^5$ & $6.1 \times 10^6$ \\
Solid particles						   & $2.0 \times 10^5$ & $8.0 \times 10^5$ & $6.7 \times 10^5$ & $2.7 \times 10^6$ \\
Simulation time						   & \multicolumn{2}{c}{$3 \: s$} & \multicolumn{2}{c}{$3 \: s$} \\
Time step ($\Delta t$) 				   & $7.0 \times 10^{-5} \: s^{a}$ & $4.0 \times 10^{-5} \: s^{a}$ & $4.0 \times 10^{-4} \: s$ & $2.0 \times 10^{-4}  \: s$ \\
\multirow{2}{9em}{Computational time}  & (CPU 0d4h00m$^b$) & (CPU 3d1h00m$^b$) & \multirow{2}{8em}{(CPU 0d5h00m$^b$)} & \multirow{2}{8em}{(CPU 6d0h00m$^b$)} \\
									   & (GPU 0d0h40m$^c$) & (GPU 0d9h30m$^c$) & & \\ \botrule
\end{tabular} }
\begin{tabfootnote}
\tabmark{a} Based on the number of computational steps. DualSPHysics uses an adaptative time step restrict by advection and diffusion stability criteria (see Eq.~(\ref{eq:sphdt}))\\
\tabmark{b} CPU Intel \textsuperscript{\textregistered} Xeon \textsuperscript{\textregistered} Processor E5 v2 Family, processor base frequency of 2.80 GHz, 20 cores and 126 GB of memory\\
\tabmark{c} GPU Nvidia \textsuperscript{\textregistered} Tesla \textsuperscript{\textregistered} K40m, 15 Multiprocessors (2880 cores), clock rate of 0.75 GHz and 11.519 GB of global memory
\end{tabfootnote}
\end{table}

\begin{figure}[th]
\centerline{\includegraphics[width=13cm]{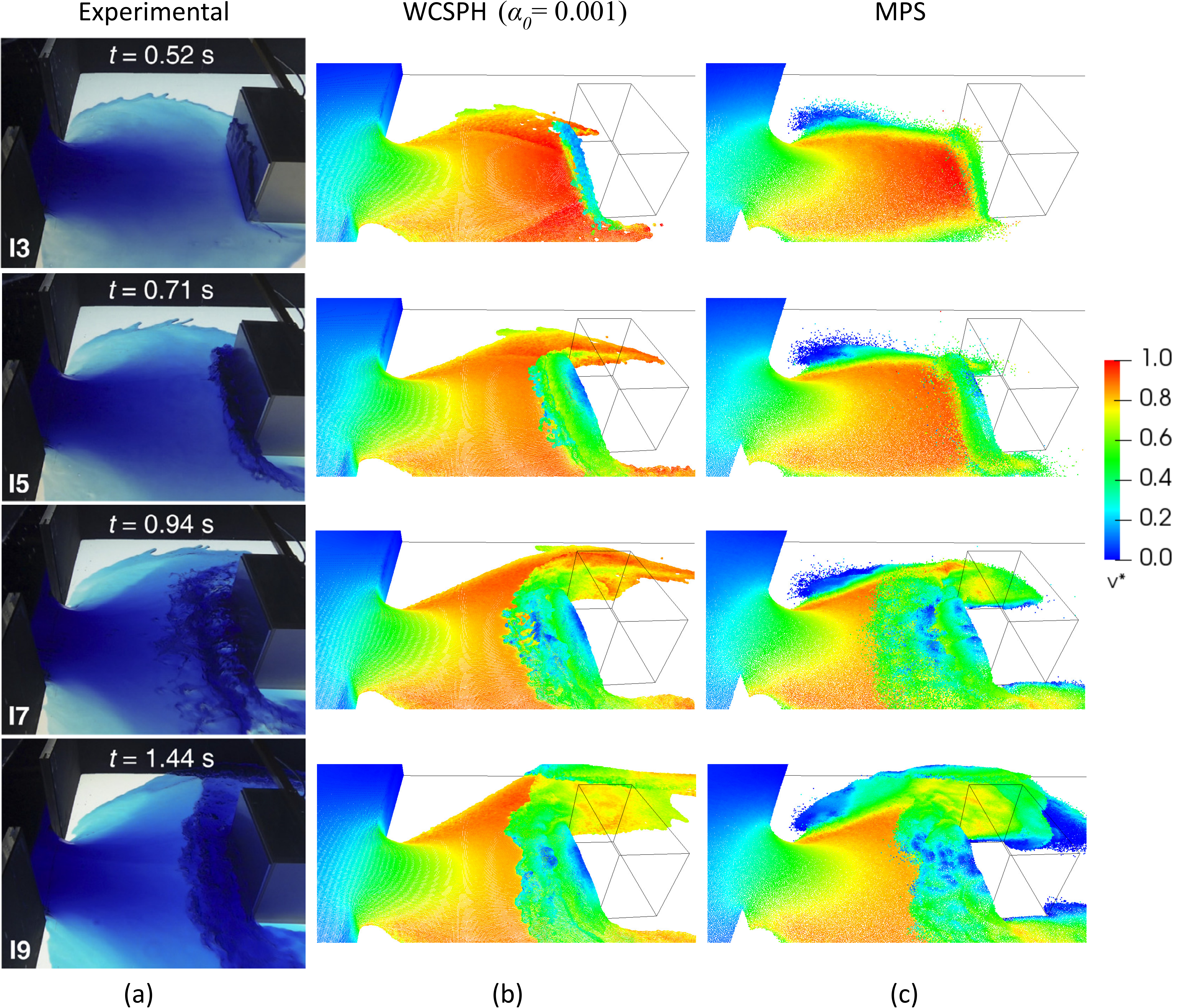}}
\vspace*{8pt}
\caption{Snapshots of the dam-break experiment (a) \protect\cite{aureli15}, and dimensionless velocity field obtained by WCSPH adopting (b) $\alpha_0 = 0.001$ and (c) MPS. Particle distance $l^0 = 0.0025 m$.\label{eleven}}
\vspace*{-\baselineskip}
\end{figure}

Comparisons of the dimensionless velocities fields $\text{v}^* = \frac{\vert \mathbf{u} \vert}{\sqrt{g H_w}}$ obtained by WCSPH adopting $\alpha_0 = 0.001$ and MPS, both using the initial particle distance $l^0 = 0.0025 m$, are shown in Fig.~\ref{eleven}. Snapshots of the experiment also are presented.

From Fig.~\ref{eleven}, the first stages of the wall impact and wave run-up on the wall are well reproduced by both methods. Nevertheless, the flow spreading downstream the gate is better reproduced by the present MPS simulation. The high-intensity velocity field near the wall computed by the WCSPH leads to the narrow-spreading flow towards downstream. As pointed out in \citeauthor{aureli15} \shortcite{aureli15}, the WCSPH is strongly affected by the artificial viscosity, which might be related to the lower flow spreading, even for the lower artificial viscosity cases simulated in the present study.

\begin{figure}[th]
\centerline{\includegraphics[width=13cm]{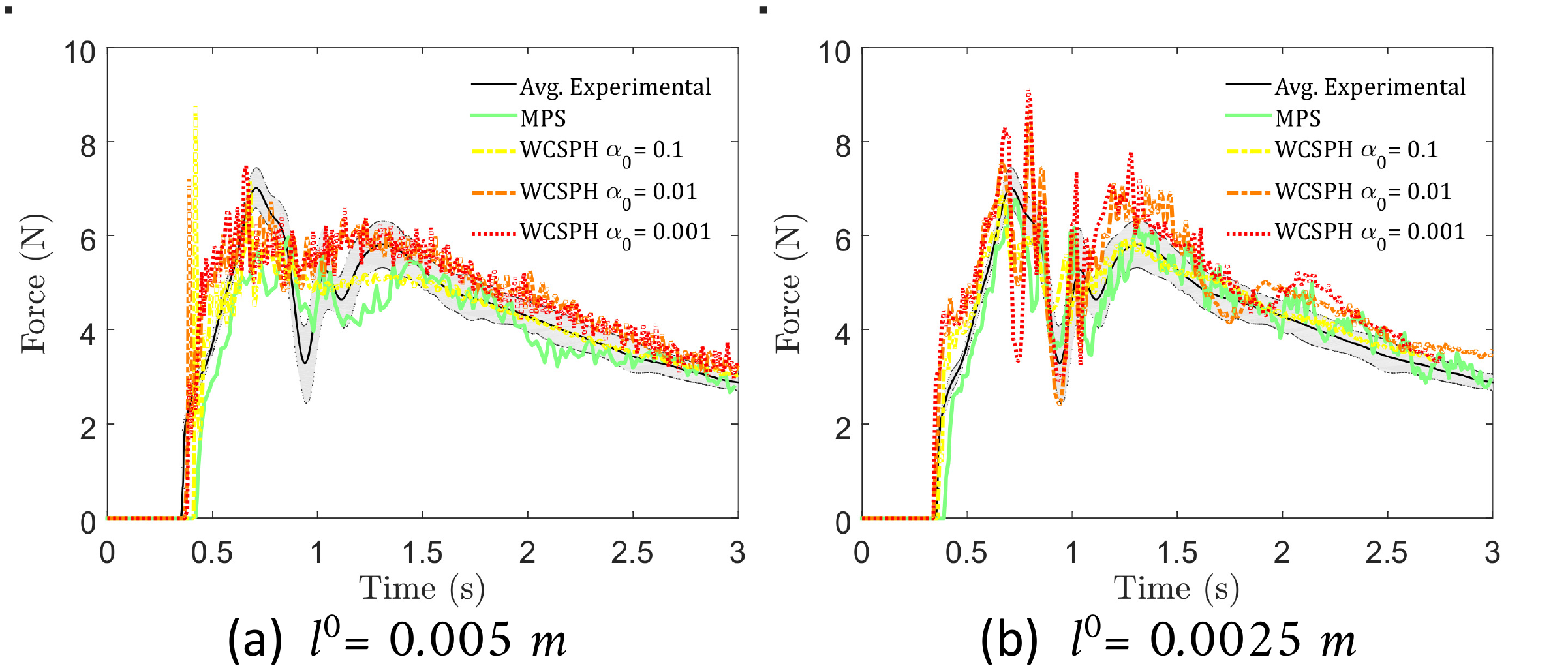}}
\vspace*{8pt}
\caption{Force time series on the block. Experimentally measured \protect\cite{aureli15} and numerically computed using MPS and WCSPH for the initial particle distance (a) $l^0 = 0.005 m$ and (b) $l^0 = 0.0025 m$.\label{twelve}}
\vspace*{-\baselineskip}
\end{figure}

The experimentally measured average force \cite{aureli15} and the numerically computed forces on the box for $l^0 = 0.005 m$ and $0.0025 m$ are presented in Figs.~\ref{twelve}(a) and \ref{twelve}(b), respectively. It is important to mention how the forces on the box weather side were obtained for each method. For the MPS, the force was calculated by the summation of the pressures of wall particles multiplied by the vertical area of the box. For the WCSPH, the default output force of DualSPHysics is obtained by the summation of the acceleration of the wall particles, see Eq.~(\ref{eq:momentum}), multiplied by the mass of the particles at the box weather side. Moreover, accelerations of the wall particles are not considered in the motions of fixed or forced solids in DualSPHysics, i.e., they are used only to obtain the force.

For the simulations with $l^0 = 0.005 m$, illustrated in Fig.~\ref{twelve}(a), the numerical wave front propagates more slowly towards the obstacle, and the initial instant of the impact is slightly delayed. The first peak force is underestimated by the MPS, while it is better reproduced by the WCSPH. After that, forces computed by MPS and WCSPH, especially when $\alpha_0 = 0.1$, i.e., $\nu_0 = \mathcal{O}(10^{-3})$, match well the experimental one.

Higher resolution models with smaller particle distance $l^0 = 0.0025 m$ improves the accuracy of the forces computed by MPS and WCSPH with $\alpha_0 = 0.1$, i.e., $\nu_0 = \mathcal{O}(10^{-3})$, as shown in Fig.~\ref{twelve}(b). For both simulations, the computed first peak force agrees very well with the experimental one. However, the decrease of the artificial viscosity leads to high-amplitude force oscillations in the WCSPH.

As in the previous case, see Table~\ref{tab1}, the CPU costs of the WCSPH simulations is lower than that demanded by MPS, especially when the number of particles is increased, with $CPUtime_{MPS} \approx 2.0 \times CPUtime_{WCSPH}$. Again, WCSPH simulations by using GPU reduce significantly the processing time in relation to the CPU-based code, given $CPUtime_{WCSPH} \approx 6.0$ to $8.0 \times GPUtime_{WCSPH}$, see Table~\ref{tab3}.

\section{Conclusions}
In this present study, the comparisons on the accuracy and computational time of the WCSPH and MPS methods are examined using two 3D dam-break cases. Based on the WCSPH results of the first case, the fluid behavior is well reproduced when $\alpha_0 \leq 0.01$, i.e., $\nu_0 \leq \mathcal{O}(10^{-3})$, whereas the pressure impulse deviations indicate that $\alpha_0 \leq 0.001$, i.e., $\nu_0 \leq \mathcal{O}(10^{-4})$, is preferable to reproduce the hydrodynamic loads. From the second case, a better approximation of the experimental force is achieved when $\alpha_0 = 0.1$, i.e., $\nu_0 = \mathcal{O}(10^{-3})$, while the amplitude force oscillations are larger and increase with the decrease of the artificial viscosity. Therefore, in order to achieve a reasonable result, the artificial viscosity $\alpha_0$ needs to be tuned for each resolution, providing the equivalent kinematic viscosity $\nu_0$ value in the range of $\mathcal{O}(10^{-3})$ to $\mathcal{O}(10^{-4})$. Moreover, we must highlight that the optimum value of the $\alpha_0$ can be calibrated in different ways, e.g., by a linear dependency of the value of the $\alpha_0$ on the Iribarren number, a dimensionless parameter used to model several effects of wave breaking relating the bottom slope angle, the deep-water wave height and wavelength, as demonstrated in \citeauthor{padova14} \shortcite{padova14} or by empirical relation of $\alpha_0$ as function of relative wave height and slope provided in \citeauthor{he20} \shortcite{he20} for the finite particle method (FPM). Therefore, the carefully calibration of the artificial viscosity coefficient $\alpha_0$ for each case is highly recommended. Overall, the present MPS reproduced more accurately the hydrodynamic loads and provide more reliable and smooth pressure distribution with reduced pressure fluctuations. In this way, improved versions, e.g., WCSPH based on a low-dissipation Riemann solver, KGC-SPH or KGF-SPH, are recommended.  According to the simulated cases in the present work, the computational time required for WCSPH in CPU is significantly lower compared to the MPS ($CPUtime_{MPS} \approx 2 \times CPUtime_{WCSPH}$), since the latter need to solve a linear system for the pressure calculation. Furthermore, the adoption of parallel processing in GPU for the WCSPH significantly reduced the computational time, compared to the conventional serial CPU solver with $CPUtime_{WCSPH} \approx 6.0$ to $8.0 \times GPUtime_{WCSPH}$. For a better understanding of the relative benefits of the methods for solving free-surface problems, further analysis through practical applications cases and improved high-order methods are considered by the authors.

\section*{Acknowledgments}
The first author would like to thank the Coordenação de Aperfeiçoamento de Pessoal de Nível Superior - Brasil (CAPES) - Finance Code 001, for the doctorate scholarship. The first and second authors are also grateful to Petrobras for financial support on the development of the MPS/TPN-USP simulation system based on MPS method. 
Besides, the authors also express their gratitude to the anonymous reviewers for their valuable suggestions and comments to improve the quality of this paper.


\begin{thebibliography}
\bibitem[\protect\citeauthoryear{Abdelrazek {\it et~al}.}{2014}]{abdelrazek14}
Abdelrazek, A. M., Kimura, I. and Shimizu, Y. (2014). Comparison between SPH and MPS methods for numerical simulations of free surface flow problems. {\it Journal of Japan Society of Civil Engineers, Ser. B1 (Hydraulic Engineering)}, {\bf 7}: 67--72.

\bibitem[\protect\citeauthoryear{Akbari}{2018}]{akbari18}
Akbari, H. (2018). Evaluation of incompressible and compressible SPH methods in modeling dam break flows. {\it International Journal of Coastal and Offshore Engineering}, {\bf 2}: 45--57.

\bibitem[\protect\citeauthoryear{Aureli {\it et~al}.}{2015}]{aureli15}
Aureli, F., {\it et al}. (2015). Experimental and numerical evaluation of the force due to the impact of a dam-break wave on a structure. {\it Advances in Water Resources}, {\bf 76}: 29--42.

\bibitem[\protect\citeauthoryear{Bakti {\it et~al}.}{2016}]{bakti16}
Bakti, F. P., {\it et al}. (2016). Comparative study of standard WC-SPH and MPS solvers for free surface academic problems. {\it International Journal of Offshore and Polar Engineering}, {\bf 26}: 235--243.

\bibitem[\protect\citeauthoryear{Chow {\it et~al}.}{2018}]{chow18}
Chow, A. D., {\it et al}. (2018). Incompressible SPH (ISPH) with fast Poisson solver on a GPU. {\it Computer Physics Communications}, {\bf 226}: 81--103.

\bibitem[\protect\citeauthoryear{Courant {\it et~al}.}{1967}]{courant67}
Courant, R., Friedrichs, K. and Levy, H. (1967). On the partial difference equations of mathematical physics. {\it IBM Journal of Research and Development}, {\bf 11}: 215--234.

\bibitem[\protect\citeauthoryear{Crespo {\it et~al}.}{2007}]{crespo07}
Crespo, A. J. C., Gomez-Gesteira, M. and Dalrymple, R. A. (2007). Boundary conditions generated by dynamic particles in SPH methods. {\it CMC: Computers, Materials, \& Continua}, {\bf 5}: 173--184.

\bibitem[\protect\citeauthoryear{Crespo {\it et~al}.}{2015}]{crespo15}
Crespo, A. J. C., {\it et al}. (2015). DualSPHysics: Open-source parallel CFD solver on smoothed particle hydrodynamics (SPH). {\it Computer Physics Communications}, {\bf 187}: 204--216.

\bibitem[\protect\citeauthoryear{Cummins and Rudman}{1999}]{cummins99}
Cummins, S. J. and Rudman, M. (1999). An SPH projection method. {\it Journal of Computational Physics}, {\bf 152}: 584--607.

\bibitem[\protect\citeauthoryear{De Padova {\it et~al}.}{2014}]{padova14}
De Padova, D., Dalrymple, R. A. and Mossa, M. (2014). Analysis of the artificial viscosity in the smoothed particle hydrodynamics modelling of regular waves. {\it Journal of Hydraulic Research}, {\bf 52}: 836--848.

\bibitem[\protect\citeauthoryear{Fernandes {\it et~al}.}{2015}]{fernandes15}
Fernandes, D. T., {\it et al}. (2015). A domain decomposition strategy for hybrid parallelization of moving particle semi-implicit (MPS) method for computer cluster. {\it Cluster Computing}, {\bf 18}: 1363--1377.

\bibitem[\protect\citeauthoryear{Flebbe {\it et~al}.}{1994}]{flebbe94}
Flebbe, O., {\it et al}. (1994). Smoothed Particle Hydrodynamics: Physical viscosity and the simulation of accretion disks. {\it The Astrophysical Journal}, {\bf 431}: 740--760.

\bibitem[\protect\citeauthoryear{Gingold and Monagham}{1977}]{gingold77}
Gingold, R. A. and Monagham, J. J. (1977). Smoothed particle hydrodynamics: Theory and application to non-spherical stars. {\it Monthly Notices of the Royal Astronomical Society}, {\bf 181}: 375--389.

\bibitem[\protect\citeauthoryear{Gomez-Gesteira {\it et~al}.}{2012}]{gesteira12}
Gomez-Gesteira, M., {\it et al}. (2012). SPHysics - development of a free-surface fluid solver - Part 1: Theory and formulations. {\it Computers \& Geosciences}, {\bf 48}: 289--299.

\bibitem[\protect\citeauthoryear{Gotoh and Khayyer}{2016}]{gotoh16}
Gotoh, H. and Khayyer, A. (2016). Current achievements and future perspectives for projection-based particle methods with applications in ocean engineering. {\it Journal of Ocean Engineering and Marine Energy}, {\bf 2}: 251--278.

\bibitem[\protect\citeauthoryear{Gotoh and Khayyer}{2018}]{gotoh18}
Gotoh, H. and Khayyer, A. (2018). On the state-of-the-art of particle methods for coastal and ocean engineering. {\it Coastal Engineering Journal}, {\bf 60}: 79--103.

\bibitem[\protect\citeauthoryear{Guo {\it et~al}.}{2018}]{guo18}
Guo, X., {\it et al}. (2018). New massively parallel scheme for incompressible smoothed particle hydrodynamics (ISPH) for highly nonlinear and distorted flow. {\it Computer Physics Communications}, {\bf 233}: 16--28.

\bibitem[\protect\citeauthoryear{Hashimoto {\it et~al}.}{2013}]{hashimoto13}
Hashimoto, H., Grenier, N. and Le Touzé, D. (2013). Comparisons of MPS and SPH methods: Forced roll test of a two-dimensional damaged car deck. {\it Proceedings of Japan Society of Naval Architecture \& Ocean Engineering}.

\bibitem[\protect\citeauthoryear{He {\it et~al}.}{2020}]{he20}
He, F., {\it et al}. (2020). Numerical investigation of the solitary wave breaking over a slope by using the finite particle method. {\it Coastal Engineering}, {\bf 156}.

\bibitem[\protect\citeauthoryear{Hori {\it et~al}.}{2011}]{hori11}
Hori, C., {\it et al}. (2011). GPU-acceleration for moving particle semi-Implicit method. {\it Computers \& Fluids}, {\bf 51}: 174--183.

\bibitem[\protect\citeauthoryear{Huang {\it et~al}.}{2019}]{huang19}
Huang, C., {\it et al}. (2019). A kernel gradient free SPH method with iterative particle shifting technology for modeling low Reynolds flows around airfoils. {\it Engineering Analysis with Boundary Elements}, {\bf 106}: 571--587.

\bibitem[\protect\citeauthoryear{Hughes and Graham}{2010}]{hughes10}
Hughes, J. P. and Graham, D. I. (2010). Comparison of incompressible and weakly-compressible SPH models for free-surface water flows. {\it Journal of Hydraulic Research}, {\bf 48}: 105--117.

\bibitem[\protect\citeauthoryear{Ikeda {\it et~al}.}{2001}]{ikeda01}
Ikeda, H., {\it et al}. (2001). Numerical analysis of jet injection behavior for fuel-coolant interaction using particle method. {\it Journal of Nuclear Science and Technology}, {\bf 38}: 174--182.

\bibitem[\protect\citeauthoryear{Jian {\it et~al}.}{2016}]{jian16}
Jian, W., {\it et al}. (2016). Smoothed particle hydrodynamics simulations of dam-break flows around movable structures. {\it International Journal of Offshore and Polar Engineering}, {\bf 26}: 33--40.

\bibitem[\protect\citeauthoryear{Kleefsman {\it et~al}.}{2005}]{kleefsman05}
Kleefsman, K. M. T., {\it et al}. (2005). A Volume-of-Fluid based simulation method for wave impact problems. {\it Journal of Computational Physics}, {\bf 206}: 363--393.

\bibitem[\protect\citeauthoryear{Kolukisa {\it et~al}.}{2017}]{kolukisa17}
Kolukisa, D. C., Ozbulut, M. and Pesman, E., (2017). An investigation on the effects of time integration schemes on weakly compressible SPH method. {\it In VII International Conference on Computational Methods in Marine Engineering MARINE 2017}.

\bibitem[\protect\citeauthoryear{Koshizuka and Oka}{1996}]{koshizuka96}
Koshizuka, S. and Oka, Y. (1996). Moving-particle semi-implicit method for fragmentation of incompressible fluid. {\it Nuclear Science and Engineering}, {\bf 123}: 421--434.

\bibitem[\protect\citeauthoryear{Koshizuka {\it et~al}.}{1999}]{koshizuka99}
Koshizuka, S., Ikeda, H. and Oka, Y. (1999). Numerical analysis of fragmentation mechanisms in vapor explosions. {\it Nuclear Engineering and Design}, {\bf 189}: 423--433.

\bibitem[\protect\citeauthoryear{Koshizuka {\it et~al}.}{2018}]{koshizuka18}
Koshizuka, S., {\it et al}. (2018). {\it Moving Particle Semi-implicit Method: A Meshfree Particle Method for Fluid Dynamics}, Academic Press.

\bibitem[\protect\citeauthoryear{Lee {\it et~al}.}{2008}]{lee08}
Lee, E.-S., {\it et al}. (2008). Comparisons of weakly compressible and truly incompressible algorithms for the SPH mesh free particle method. {\it Journal of Computational Physics}, {\bf 227}: 8417--8436.

\bibitem[\protect\citeauthoryear{Liu and Liu}{2015}]{liu15}
Liu, M. B. and Liu, G. R. (2015). {\it Particle Methods for Multi-Scale and Multi-Physics}, World Scientific Publishing Company, Singapore.

\bibitem[\protect\citeauthoryear{Liu and Zhang}{2019}]{liu19}
Liu, M. and Zhang, Z. (2019). Smoothed particle hydrodynamics (SPH) for modeling fluid-structure interactions. {\it Science China Physics, Mechanics \& Astronomy}, {\bf 62}.

\bibitem[\protect\citeauthoryear{Lucy}{1977}]{lucy77}
Lucy, L. B. (1977). A numerical approach to the testing of the fission hypothesis. {\it Astronomical Journal}, {\bf 82}: 1013--1024.

\bibitem[\protect\citeauthoryear{Mabssout and Herreros}{2013}]{mabssout13}
Mabssout, M. and Herreros, M. I. (2013). Runge–Kutta vs Taylor-SPH: Two time integration schemes for SPH with application to soil dynamics. {\it Applied Mathematical Modelling}, {\bf 37}: 3541--3563.

\bibitem[\protect\citeauthoryear{Molteni and Colagrossi}{2009}]{molteni09}
Molteni, D. and Colagrossi, A. (2009). A simple procedure to improve the pressure evaluation in hydrodynamic context using the SPH. {\it Computer Physics Communications}, {\bf 180}: 861--872.

\bibitem[\protect\citeauthoryear{Monaghan}{1992}]{monaghan92}
Monaghan, J. J. (1992). Smoothed particle hydrodynamics. {\it Annual Review of Astronomy and Astrophysics}, {\bf 30}: 543--574.

\bibitem[\protect\citeauthoryear{Monaghan}{1994}]{monaghan94}
Monaghan, J. J. (1994). Simulating free surface flows with SPH. {\it Journal of Computational Physics}, {\bf 110}: 399--406.

\bibitem[\protect\citeauthoryear{Monaghan and Kos}{1999}]{monaghan99}
Monaghan, J. J. and Kos, A. (1999). Solitary waves on a Cretan beach. {\it Journal of Waterway, Port, Coastal, and Ocean Engineering}, {\bf 125}: 145--154.

\bibitem[\protect\citeauthoryear{Monaghan}{2005}]{monaghan05}
Monaghan, J. J. (2005). Smoothed particle hydrodynamics. {\it Reports on Progress in Physics}, {\bf 68}: 1703--1759.

\bibitem[\protect\citeauthoryear{Moulinec {\it et~al}.}{2008}]{moulinec08}
Moulinec, C., {\it et al}. (2008). Parallel 3-D SPH simulations. {\it Computer Modeling in Engineering and Sciences}, {\bf 25}: 133--148.

\bibitem[\protect\citeauthoryear{Rafiee {\it et~al}.}{2012}]{rafiee12}
Rafiee, A., {\it et al}. (2012). Comparative study on the accuracy and stability of SPH schemes in simulating energetic free-surface flows. {\it European Journal of Mechanics - B/Fluids}, {\bf 36}: 1--16.

\bibitem[\protect\citeauthoryear{Shao {\it et~al}.}{2012}]{shao12}
Shao, J. R., {\it et al}. (2012). An improved SPH method for modeling liquid sloshing dynamics. {\it Computers \& Structures}, {\bf 100-101}: 18--26.

\bibitem[\protect\citeauthoryear{Shao {\it et~al}.}{2016}]{shao16}
Shao, J. R., Li, S. M. and Liu, M. B. (2016). Numerical simulation of violent impinging jet flows with improved SPH method. {\it International Journal of Computational Methods}, {\bf 13}.

\bibitem[\protect\citeauthoryear{Shimizu and Gotoh}{2016}]{shimizu16}
Shimizu, Y. and Gotoh, H. (2016). Toward enhancement of MPS method for ocean engineering: Effect of time-integration schemes. {\it International Journal of Offshore and Polar Engineering}, {\bf 26}: 374--384.

\bibitem[\protect\citeauthoryear{Souto-Iglesias {\it et~al}.}{2013}]{souto13}
Souto-Iglesias, A., {\it et al}. (2013). On the consistency of MPS. {\it Computer Physics Communications}, {\bf 184}: 732--745.

\bibitem[\protect\citeauthoryear{Tsukamoto {\it et~al}.}{2016}]{tsukamoto16}
Tsukamoto, M. M., Cheng, L.-Y. and Motezuki, F. K. (2016). Fluid interface detection technique based on neighborhood particles centroid deviation (NPCD) for particle methods. {\it International Journal for Numerical Methods in Fluids}, {\bf 82}: 148--168.

\bibitem[\protect\citeauthoryear{Tsukamoto {\it et~al}.}{2020}]{tsukamoto20}
Tsukamoto, M. M., {\it et al}. (2020). A numerical study of the effects of bottom and sidewall stiffeners on sloshing behavior considering roll resonant motion. {\it Marine Structures}, {\bf 72}.

\bibitem[\protect\citeauthoryear{Vaz {\it et~al}.}{2009}]{vaz09}
Vaz, G., Jaouen, F. and Hoekstra, M. (2009). Free-surface viscous flow computations: Validation of URANS code FreSCo. {\it Proceedings of the 28th International Conference on Ocean, Offshore and Arctic Engineering}.

\bibitem[\protect\citeauthoryear{Verlet}{1967}]{verlet67}
Verlet, L. (1967). Computer experiments on classical fluids. I. Thermodynamical properties of Lennard-Jones molecules. {\it Physical Review}, {\bf 159}: 98--103.

\bibitem[\protect\citeauthoryear{Violeau and Leroy}{2014}]{violeau14}
Violeau, D. and Leroy, A. (2014). On the maximum time step in weakly compressible SPH. {\it Journal of Computational Physics}, {\bf 256}: 388--415.

\bibitem[\protect\citeauthoryear{Wang {\it et~al}.}{2017}]{wang17}
Wang, L., Jiang, Q. and Zhang, C. (2017). Improvement of moving particle semi-implicit method for simulation of progressive water waves. {\it International Journal for Numerical Methods in Fluids}, {\bf 85}: 69--89.

\bibitem[\protect\citeauthoryear{Wendland}{1995}]{wendland95}
Wendland, H. (1995). Piecewise polynomial, positive definite and compactly supported radial functions of minimal degree. {\it Advances in Computational Mathematics}, {\bf 4}: 389--396.

\bibitem[\protect\citeauthoryear{Wong {\it et~al}.}{2019}]{ye19}
Ye, T., {\it et al}. (2019). Smoothed particle hydrodynamics (SPH) for complex fluid flows: Recent developments in methodology and applications. {\it Physics of Fluids}, {\bf 31}.

\bibitem[\protect\citeauthoryear{Zhan {\it et~al}.}{2017}]{zhang17}
Zhang, C., Hu, X. Y. and Adams, N. A. (2017). A weakly compressible SPH method based on a low-dissipation Riemann solver. {\it Journal of Computational Physics}, {\bf 335}: 605--620.

\bibitem[\protect\citeauthoryear{Zheng {\it et~al}.}{2014}]{zheng14}
Zheng, X., Ma, Q.-W. and Duan, W.-Y. (2014). Comparative study of different SPH schemes on simulating violent water wave impact flows. {\it China Ocean Engineering}, {\bf 28}: 791--806.

\end{thebibliography}
\end{document}